\documentclass[useAMS,usenatbib]{mn2e}
\usepackage{graphicx}
\usepackage{bm}
\usepackage{amsmath}

\newcommand{\be}{\begin{equation}}
\newcommand{\ee}{\end{equation}}

\title[Non-resonant streaming instability near  magnetized relativistic shocks]{Non-resonant magnetohydrodynamics streaming instability near magnetized relativistic shocks}  
\author[F. Casse, A. Marcowith and R. Keppens ]{F. Casse $^{1}$\thanks{fcasse@apc.univ-paris7.fr}, A. Marcowith$^{2}$ and R. Keppens$^{3}$\\
$^{1}$ Laboratoire AstroParticule \& Cosmologie (APC), Universit\'e Paris Diderot, CNRS/IN2P3, CEA/Irfu, Observatoire de Paris,\\ Sorbonne Paris Cit\'e  -   \ 10, rue Alice Domon et L\'eonie Duquet, F-75205 Paris Cedex 13, France\\
$^{2}$ Laboratoire Univers et Particules (LUPM) Universit\'e Montpellier, CNRS/IN2P3, CC72, place Eug\`ene Bataillon,  \\ F-34095 Montpellier Cedex 5, France\\
$^{3}$Centre for mathematical Plasma Astrophysics, Department of Mathematics, KU Leuven, Celestijnenlaan 200B, 3001 Heverlee, Belgium\\
}
\begin{document}

\date{}

\maketitle

\begin{abstract}
We present in this paper both a linear study and numerical relativistic MHD simulations of the non-resonant streaming instability occurring in the precursor of relativistic shocks. In the shock front restframe, we perform a linear analysis of this instability in a likely configuration for ultra-relativistic shock precursors. This considers magneto-acoustic waves having a wave vector perpendicular to the shock front and the large scale magnetic field. Our linear analysis is achieved without any assumption on the shock velocity and is thus valid for all velocity regimes. In order to check our calculation, we also perform relativistic MHD simulations describing the propagation of the aforementioned magneto-acoustic waves through the shock precursor. The numerical calculations confirm our linear analysis, which predicts that the growth rate of the instability is maximal for ultra-relativistic shocks and exhibits a wavenumber dependence $\propto k_x^{1/2}$. Our numerical simulations also depict the saturation regime of the instability where we show that the magnetic amplification is moderate but nevertheless significant ($\delta B/B\leq 10$). This latter fact may explain the presence of strong turbulence in the vicinity of relativistic magnetized shocks. Our numerical approach also introduces a convenient means to handle isothermal (ultra-)relativistic MHD conditions.
\end{abstract}


\section{Introduction}
Astrophysical shock waves and particle acceleration up to the highest energies are intimately connected processes. In non-relativistic supernova remnants a large amplification of the magnetic field at the forward shock has been detected by recent high angular resolution observations from X-ray satellites Chandra and XMM Newton \citep{Ballet06}. The very nature of the amplification process has been the object of intense research work in the last decade (see a recent review by \citet{Schure+12}).
It appeared that the current generated by the energetic particles (hereafter abusively considered as cosmic rays) streaming ahead of the shock front in the so-called CR precursor could provide such an amplification \citep{Bell04} even if the definite value of the amplified magnetic field remains uncertain \citep{Riquelme09}. In relativistic shocks, notably in gamma-ray bursts we can only indirectly evaluate magnetic field amplification in the shock precursor. Here again high magnetic amplification seems to be required in the upstream medium \citep{Li06}. The fraction $\epsilon_B$ of the shock energy flux imparted into the magnetic field has been inferred from multi-wavelength observations of gamma-ray burst afterglows and can reach values up to $1\%$ (however see the discussion in \citet{Lemoine13}). Here again the streaming of cosmic rays (CRs) ahead of the shock may provide the main source of free energy to generate the magnetic field \citep{Milos06}. \\

In a previous work \citet{Pelletier+09} (hereafter PLM09) have investigated in detail the amplification of magnetic disturbances at relativistic shock waves. This results in two necessary conditions to get the Fermi acceleration process  working in such environments. The first condition states that the turbulence triggered around the shock must be at small scale (with respect to the triggering particle Larmor radius). The second condition states that the ratio of the perturbed to background magnetic field $\delta B/ B_0 \gg 1$ consistent with strong magnetic field amplification. PLM09 did perform 1 and 2D analytical analysis of magnetohydrodynamic (MHD) waves evolution in the CR precursor of highly relativistic perpendicular shocks. They have shown that Alfv\'en waves are stable and that magneto-acoustic waves can be destabilized by the charge of CR ahead the shock. Hence it has been further shown that relativistic shocks can trigger microscopic instabilities that lead to turbulence at scales beyond the MHD regime \citep{Lemoine10, Bret09}, one may also see the recent review by \citet{Bykov+12}. However, MHD waves are of great interest because they can provide a better confinement of high energy particles. They also can at least partly explain values of $\epsilon_B$ inferred from observations. 
Finally they can also generate shock trains upstream the shock front that can contribute to particle acceleration and plasma heating. In this work we continue the analysis initiated by PLM09 by testing these analytical results with respect to relativistic MHD (RMHD) numerical simulations and by providing a more general linear analytical framework that can be generalized to more complex flow configurations.  In this work we also provide an analysis of mildly- and non-relativistic flow regimes.\\

The paper is organized as follows. In section \ref{Linear} we discuss the physical case considered in this work as well as the formalism used in the linear instability analysis. In section \ref{RMHD} we give a short presentation of the relativistic version of MPI-AMRVAC and hence describe the different stages of the streaming instability in a reference case corresponding to an ultra-relativistic shock wave. We end this section with a parametric survey of the instability with respect to the shock Lorentz factor and to the perturbation wavenumber. We conclude and provide a short discussion in section \ref{Summary}.

\section{ Non-resonant streaming instability near relativistic shocks}
\label{Linear}
As pointed out by PLM09, ultra-relativistic astrophysical shock waves are likely to exhibit a perpendicular large scale magnetic structure, which means a magnetic field parallel to the shock front (hence perpendicular to the shock normal). Indeed, the Lorentz transformation of a magnetic field from the observer rest-frame (denoted with "obs" subscripts) to the shock rest-frame (denoted as "sh" subscripts) leads to 
\begin{eqnarray}
B_{\parallel, \rm sh} &=& B_{\parallel, \rm obs} \nonumber \\
{\bf B}_{\perp,\rm sh} &=& \Gamma_{\rm sh}\left({\bf B}_{\perp,\rm obs} - \bm{\beta}_{\rm sh}\times {\bf E}_{\perp,\rm obs}\right)
\end{eqnarray}
where $\bm{\beta}_{\rm sh}$ is the velocity of the shock wave in the observer frame ($\Gamma_{\rm sh}^{-2} = 1 - \beta^2_{\rm sh}$) while ${\bf E}_{\perp,\rm obs}$ is the electromotive field occurring in the observer frame. Ultra-relativistic shock waves do have generically a perpendicular magnetic field in the frame of the shock, except if the magnetic field in the observer frame is aligned with the normal direction to the shock front within ${\cal O}(1/\Gamma_{\rm sh})$. This work considers hereafter the case of a pure perpendicular magnetic field in the shock frame where the magnetic field is oriented in a $y$ direction parallel to the shock front. The shock front will be assumed as planar thus leading to fluid properties varying only along the normal to the shock front in the $x$ direction. 
\subsection{Special relativistic magnetohydrodynamics equations}
\label{Sect:RMHD}
In this work we present time dependent numerical simulations of the evolution of MHD waves that propagate through the precursor of a perpendicular magnetized shock wave. The special relativistic magnetohydrodynamic (SRMHD) equations ruling both the fluid and magnetic field evolution are displayed in a conservative form in order to ease the comparison between analytical and numerical calculations. We have 
\begin{eqnarray}
D_tD + {\bf \nabla}\cdot\left(D\bm{\beta}\right) = 0 &&\\
D_t{\bf S} + {\bf \nabla}\cdot\left({\bf S}\bm{\beta} -\left(\frac{\bf B}{\Gamma^2}+(\bm{\beta}\cdot {\bf B})\bm{\beta}\right)\frac{{\bf B}}{4\pi} + P_{\rm tot}\bf{I}\right) = 0 && \\
D_t{\bf B} + \nabla\cdot\left(\bm{\beta} {\bf B} - {\bf B}\bm{\beta}\right) = 0 && \\
D_t\tau + \nabla\cdot\left( (\tau + P_{\rm tot})\bm{\beta} - (\bm{\beta}\cdot{\bf B})\frac{{\bf B}}{4\pi}\right) = 0 && \\
\nabla\cdot{\bf B} = 0 &&
\end{eqnarray}    
where the conservative variables are the relativistic mass density $D=\rho\Gamma$, the relativistic momentum ${\bf S} =  (\Gamma^2\rho h+B^2)\bm{\beta} - (\bm{\beta}\cdot {\bf B}){\bf B}$, the magnetic field ${\bf B}$ and the total energy $ \tau = \Gamma^2\rho h+B^2/4\pi- P_{\rm tot}$ respectively.  The velocity (normalized to the speed of light $c$) is denoted by $\bm{\beta}$ while $\Gamma^{-2}=1-\beta^2$ is the fluid Lorentz factor. The quantity $\rho h$ stands for the enthalpy $\rho h=\rho c^2+\gamma P/(\gamma-1)$ and $P_{\rm tot}=(B^2/\Gamma^2+(\bm{\beta}\cdot{\bf B})^2)/8\pi+P$ is the total pressure. The derivative operator $D_t=\partial_t/c$. This set of equations is the mathematical translation of mass, momentum and energy conservation while the last relation ensures that the magnetic field is monopole-free. The characteristic speeds of magneto-acoustic and Alfv\'en waves are derived by linearizing the previous equations. In the 1D configuration adopted hereafter the Alfv\'en disturbance cannot propagate along the normal to the shock front since we assume the magnetic field to be parallel to the shock front. We will hence consider only magneto-acoustic disturbances even if the case of an Alfv\'en disturbance is shortly discussed at the end of section \ref{Linanaly} (basically to show that Alfv\'en waves are stable in this flow configuration). The general slow and fast magneto-acoustic characteristic speeds $\lambda_{S,F}$ in SRMHD are obtained by solving a cumbersome quartic polynomial as discussed in \citet{vdHol08} and for which the group speed anisotropies are shown more clearly in \citet{kepp08}. In the case of a relativistic shock having a magnetic field perpendicular to the shock normal, 
the magnetic energy density and thermal pressure are very small compared to the rest mass energy density. This leads to a simpler quartic polynomial which provides the expression of $\lambda_{S,F}$
\be
\frac{\lambda_{S}}{c} \simeq \beta_x \ {\rm and } \ \frac{\lambda_{F}}{c} = \beta_x + \frac{1}{\Gamma^2}\left(\frac{c^2_S}{c^2} + \frac{B^2}{4\pi\Gamma^2\rho h}\right)^{1/2}\frac{k_x}{|k_x|}
\label{Eq:propagation}
\ee
where the sound speed is $\beta_s=c_S/c=\sqrt{\gamma P/\rho h}$ and the Alfv\'en speed is $\beta_a=B_y/\Gamma\sqrt{4\pi\rho h}$. A fast magneto-acoustic wave propagating in the upstream medium and propagating towards the shock will then display perturbed quantities as:
\be
\frac{\delta B}{B} = \frac{\delta \rho}{\rho} = \frac{k_x}{|k_x|}\frac{\Gamma^2\delta \beta_x}{\left(\displaystyle\frac{c_S^2}{c^2}+\frac{B^2}{4\pi\Gamma^2\rho h}\right)^{1/2}}=\frac{k_x}{|k_x|}\frac{\Gamma^2\delta \beta_x}{\beta_F}
\label{Initperturb}
\ee
where $k_x$ stands for the wave vector of the magneto-acoustic wave while $\beta_F= \lambda_F/c$. 

To complete this study we have to account for the presence of cosmic rays (CRs) in the shock precursor in the previous set of equations. These supra-thermal particles carry electric charges that modify the Maxwell-Gauss equation (see also PLM09):
 \be
 \bm{\nabla}\cdot\bm{E} = 4\pi\left(\rho_{\rm pl} + \rho_{\rm CR}\right)
 \ee
where $\rho_{\rm pl}$ stands for the thermal plasma charge density while $\rho_{\rm CR}$ is the charge density of the CRs. The electric field is related to the velocity and magnetic field through Ohm's law stating that in ideal MHD we have
\be
\bf{E} = -\bm{\beta}\times\bf{B}
\ee
As a result the thermal plasma composing the precursor region is not neutral and we then have to consider the cosmic ray charge density in the momentum conservation equation, namely 
\be
\label{Eq:EMot}
 D_t{\bf S} + {\bf \nabla}\cdot\left({\bf S}\bm{\beta} -\frac{\bf BB}{4\pi\Gamma^2} + P_{\rm tot}\bf{I}\right) = \rho_{\rm CR}\bm{\beta}\times\bf{B} \ .
\ee
As the shock is perpendicular we have considered that ${\bm{\beta}}\cdot{\bf{B}}=0$. The electric force arises from the presence of CRs within the precursor and stands as an external force applied onto the thermal plasma. This force will then be treated as a source term in the numerical simulation.  

\subsection{Structure of the precursor of a magnetized perpendicular relativistic shock wave}
\label{Loreconst}
The stationary Maxwell equations driving the electromagnetic field in the shock frame read 
\begin{eqnarray}
\label{Eq:Source}
{\bf E}&=& \beta_z B_y {\bf e}_x - \beta_x B_y {\bf e}_z \ ,\nonumber \\
{\bf J} &=& {c \over 4\pi} \partial_x B_y  {\bf e}_z \ , \nonumber \\ 
\rho_{\rm pl} &=& -\rho_{\rm CR} + {1 \over 4\pi} \partial_x E_x \ .
\end{eqnarray}
In the steady Amp\`ere equation above a CR current $J_{\rm CR}$ may be added to the right hand side. If such a current is taken to be parallel to the mean magnetic field, the equilibrium equation of motion is not modified with respect to Eq. \ref{Eq:EMot} as such an aligned current does not contribute to the Lorentz force. But CR current configurations along the shock normal would require a full 2D approach, which is beyond the scope of this work and will be considered elsewhere. 

The Faraday equation reads
\begin{equation}
\label{Eq:Faraday}
\partial_x \left(\beta_x B_y\right) = 0
\end{equation}
The steady mass conservation equation links the density to the velocity along the normal to the shock front as
\be
\partial_x\left(\Gamma\rho \beta_x\right) = 0
\ee
The stationary relativistic equation of motion in its primitive form is (see e.g. \cite{hey03})
\begin{equation}
\label{Eq:Motion}
\rho \bm{u}\cdot\bm{\nabla} {h\bm{u}} = \rho_{\rm pl} {\bf E} + {1 \over c} {\bf J} \times {\bf B} - {\bf \nabla} P 
\end{equation}
where $\bm{u}=\Gamma\bm{\beta}$ is the spatial part of the relativistic four-vector velocity, such that $\Gamma=(1+u_x^2+u_y^2+u_z^2)^{1/2}$.  Let us remind here that $h$ is defined as
\be
h= c^2+\frac{\gamma P}{(\gamma -1)\rho} = \frac{(\gamma-1)c^2}{\gamma-1-\beta^2_s}
\ee
The various components of the  equation of motion write as
\begin{eqnarray}
\label{Eq:Stat}
-\rho u_x\partial_x hu_x &-& {B_y \over 4\pi \Gamma} u_y\partial_x \frac{u_yB_y}{\Gamma} \nonumber \\
&=& \rho_{\rm CR} {u_z B_y \over \Gamma} + \partial_x\left(\frac{B_y^2}{8\pi\Gamma^2}+P\right) \nonumber \\
-\rho u_x\partial_x hu_y &=& 0  \\
\rho u_x\partial_x hu_z  &+&\frac{u_xB_y^2}{4\pi\Gamma^2}\partial_xu_z-\frac{B_y^2u_z}{4\pi\Gamma^2}\partial_xu_x= -\rho_{\rm CR} {u_x B_y \over \Gamma} \nonumber
\end{eqnarray}
Let us note that we added with respect to PLM09 the possibility to have a motion $u_y$ along the mean magnetic field, but it is obvious that $u_y$ does not depend on $x$ and can be set to $0$ in the rest of the discussion. The usual physical conditions prevailing in the interstellar medium are such that $\rho c^2\gg P\geq B_y^2/4\pi \Gamma_{\rm sh}^2$. This leads to $\beta_s\ll 1$, $\beta_F\ll 1$ and $h\simeq c^2$. Under these circumstances, we can infer the balance of the background fluid by retaining only the zeroth order terms in the previous equations, namely 
\begin{eqnarray}
\label{Eq:Stat2}
\rho c^2u_x\partial_x u_x  &=& -\rho_{\rm CR} {u_z B_y \over \Gamma}  \nonumber \\
\rho c^2u_x\partial_x u_z   &=& \rho_{\rm CR} {u_x B_y \over \Gamma} 
\label{Eq:coldequi}
\end{eqnarray}
where we easily see that $\partial_x (u_x^2+u_z^2+u_y^2) =0$ thus having a constant Lorentz factor $\Gamma=\Gamma_{\rm sh}$ throughout the precursor. This property enables us to set a direct link between the mass density, the magnetic field and the velocity $\beta_x$. Indeed, taking into account the previous equations, one can easily obtain that $\rho \beta_x$ and $\beta_xB_y$ must remain constant along the normal to the shock front and hence that $B_y/\rho$ remains also constant. 

The velocity of the fluid occurring along the $z$ direction is a direct consequence of the presence of charged CRs. The induced electromotive field is then balanced by this transverse motion of the flow. In order to assess the relative amplitude of this motion, we now consider all terms appearing in the $z$-component of the equation of motion. The constant Lorentz factor $\Gamma$ provides $u_z\partial_xu_z = -u_x\partial_xu_x$ which leads to 
\be
\left(\rho c^2u_x + \frac{B_y^2}{4\pi u_x}\frac{\Gamma_{\rm sh}^2-1}{\Gamma_{\rm sh}^2}\right)\partial_xu_z = \rho_{\rm CR}\frac{u_xB_y}{\Gamma_{\rm sh}}
 \ee 
 It is then easy to assess the transverse motion of the plasma as 
\begin{equation}
\label{Eq:Profile}
u_z(x) = \int_{x_*}^x \displaystyle\frac{\rho_{\rm CR}(x)B_y(x)}{\Gamma_{\rm sh}\rho c^2\left(1+\displaystyle\frac{B^2_y(x)(\Gamma_{\rm sh}^2-1)}{4\pi\rho(x)c^2 \Gamma_{\rm sh}^4\beta^2_x(x)}\right)}dx
\end{equation}
where $x_*$ is such that $\rho_{\rm CR}(x>x_*)=0$ (the shock front is located at $x=0$). 
The last term appearing in the denominator is proportional to the inverse of the squared Alfv\'enic Mach number measured in the observer frame. Since only strong shock waves are relevant for CR acceleration (thus super-Alfv\'enic shocks),  we may safely neglect this term. We then end up with the transverse velocity profile defined as 
\be
u_z(x) \simeq \frac{B_y(x)}{\Gamma_{\rm sh}\rho(x) c^2}\int_{x_*}^x \rho_{\rm CR}(x)dx
\label{Eq:vertequi}
\ee 
which is dimensionless since $\rho_{\rm CR} \equiv [B/L]$ ($L$ being the reference length unit of the system).
Following PLM09, we identify this length $L$ as the typical relaxation length of the CR charge density  $\ell_{\rm CR}\equiv x_*$ 
which leads to
\begin{equation}
\label{Eq:Profile-app}
|u_z|(x=0) \simeq \ell_{\rm CR} \rho_{\rm CR}(x=0){B_y \over \Gamma_{\rm sh} \rho c^2} 
\end{equation}
 We define the energy density of CR in the shock restframe as $e_{\rm CR} = \xi_{\rm CR} (\Gamma_{\rm sh}-1)\Gamma_{\rm sh} \rho_u c^2$ and the charge density as $\rho_{\rm CR} = q e_{\rm CR}/ E_*$.  In the previous expression, $E_*$  is the mean energy of the CR having their Larmor radius of the order of $\ell_{\rm CR}$ while $q$ stands for the charge of a proton ($E_*=\ell_{\rm CR}qB_y$).  Previous relations lead to an estimate of the maximal transverse velocity, 
\begin{equation}
\label{Eq:uz}
|u_z| (x=0)\simeq \xi_{\rm CR}(\Gamma_{\rm sh}-1)
\end{equation}
where $\xi_{\rm CR}$ is the fraction of the shock energy that is transferred into the CR population by Fermi acceleration. The vertical velocity profile can now be expressed as
\be
\chi(x)=\frac{\beta_z(x)}{\beta_x(0)} = \xi_{\rm CR}\frac{(\Gamma_{\rm sh}-1)}{\beta_{\rm sh}\Gamma_{\rm sh}}\frac{\beta_z(x)}{\beta_z(0)}
\ee
The derivative of the vertical velocity enables us to define a fiducial wave vector $k_*$, namely
\be
\label{Eq:kstar}
k_*(x)=\frac{\partial_xu_z}{\Gamma_{\rm sh}} = \xi_{\rm CR}\frac{(\Gamma_{\rm sh}-1)}{\ell_{\rm CR}\Gamma_{\rm sh}}\frac{\rho_{\rm CR}(x)}{\rho_{\rm CR}(0)}
\ee
In the case of ultra-relativistic shock waves, we recover the result by PLM09, i.e. $\chi(0)\simeq \xi_{\rm CR}< 1$. For mildly relativity shocks, this assessment holds while for non-relativistic shocks $\chi(0) \sim \xi_{\rm CR}\beta_{\rm sh}/2$. The equilibrium solution used in this work is displayed in figure \ref{F1}.

The gradient of the magnetic field along the precursor induces a drift velocity$\bm{\beta}_d$ and current $\bm{J}_d$ of CRs parallel to the shock front and perpendicular to $B_y$ and thus a force $\bm{J}_d\times \bm{B}_y \propto (\bm{B} \times \bm{\nabla} B) \times \bm{B}$ in the $x$ direction. Using Eq.\ref{Eq:Faraday}, Eq.\ref{Eq:kstar} and the conservation of the Lorentz factor of the flow one can deduce that $|\beta_d| \simeq \xi_{CR}^2 \ll |\beta_z|$.
Hence we find the force produced by the CRs charge to be dominant over the CR drift induced force which will not be retained hereafter. It should be noted however that the CR gradient drift is a potential interesting effect to account for, once the magnetic field is modified in the precursor. {\bf The drift current associated with the shock magnetic compression is difficult to infer since as already shown by \citet{Bege90}, the average number of CR shock front crossing is dropping for relativistic superluminal shocks to a value close to unity. The actual amplitude of the CR shock front drift current is then still an open issue for relativistic shocks so we discarded its effect in our calculation.} The magnetic field amplification is investigated in the next section and the impact of the drift is postponed to a forthcoming work.

\subsection{Linear study of the streaming  instability near perpendicular relativistic shocks}
\label{Linanaly}
\begin{figure*}
\centering
 \includegraphics[width=\textwidth]{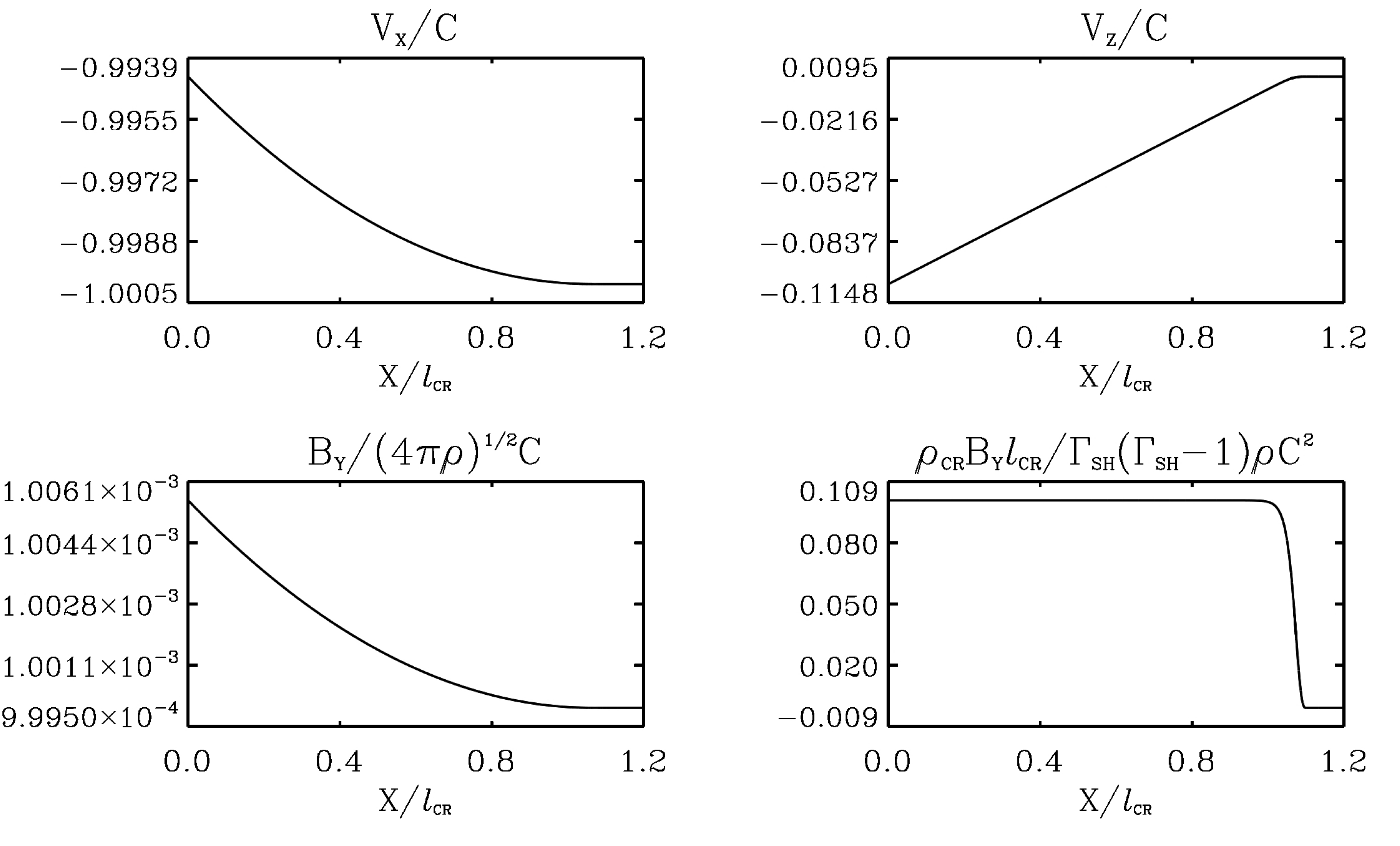}
 \caption{Initial equilibrium state of the upstream medium of an ultra-relativistic perpendicular shock ($\Gamma_{\rm sh}=100$) with a CR electrical charge. In this equilibrium, the CR energy stands for $\xi_{\rm CR}=10\%$ of the shock kinetic energy. As can be seen on the figure, the presence of the CR charge induces a parallel motion to the shock front ($v_z$) in order to balance the electrical force acting on the plasma. On the plots, the shock front is located at $x=0$ (outside the computational domain) and the unperturbed upstream region lies beyond $x_*=1.1\ell_{\rm CR}$. Our simulations consist in injecting a magneto-acoustic wave near $x_*$ and computing its propagation through the precursor from the right to the left of the box. } 
\label{F1}
\end{figure*}

The presence of a charge layer near the front shock will induce a supplementary electric force while a MHD wave is propagating through the layer. This force tends to amplify the associated velocity disturbances that will amplify the magnetic disturbances through the magnetic induction process and so on. The system then enters an exponential amplification that ultimately leads to a non-linear instability stage. PLM09 already made a first attempt to study the linear stage of the instability but discarded terms associated with the CR flow. We present in this section an alternative presentation of the linear analysis of the instability while retaining the most significant terms leading to a larger growth rate of the instability. Indeed, as we show further in this section, PLM09 has discarded terms proportional to the transverse velocity $\beta_z$ arguing that its relative amplitude is very small. As a matter of fact,  estimates achieved in (non-relativistic) supernovae shock waves have shown that the parameter $\xi_{\rm CR}$ can reach values up to $10\%$  \citep{Milos06}. According to the shock equilibrium (see figure \ref{F1}) displayed in the previous section, this leads to a maximal transverse velocity of the order of $c/10$ in relativistic shocks which cannot be discarded.\\
The basic assumptions made in this linear study is first to assume that both sonic ($\beta_s$) and Alfv\'enic speed ($\beta_a$) are much smaller that the velocity of light. The second assumption is to consider perpendicular shocks, a natural shock configuration arising from the Lorentz transformation in relativistic shocks. The first assumption is induced by the physical conditions prevailing in the interstellar medium where $\rho c^2\gg B^2/4\pi\Gamma_{\rm sh}^2 \geq P$. The second assumption enables us to assess that the wave vector of any MHD wave propagating in the precursor of a relativistic shock is mostly oriented along the normal to the shock front.  This statement arises from the Lorentz transformation inducing a wavelength contraction in the $x$ direction by a factor $\Gamma_{\rm sh}$. We use then that $k_x\gg k_*=\partial_xu_z/\Gamma_{\rm sh}$ while in this work we set $k_y=k_z=0$. \\
In order to follow the temporal evolution of MHD waves entering the precursor region, we first write the set of conservative RMHD equations describing the physics of the precursor, namely
\begin{eqnarray}
D_t(\Gamma\rho) + \partial_x (\Gamma\rho\beta_x) &=& 0  \\
D_t(\Gamma^2\rho c^2\beta_x) + \partial_x\left(\Gamma^2\rho c^2\beta_x^2 \right) &=& -\rho_{\rm CR}\beta_zB_y  \\
D_t(\Gamma^2\rho c^2\beta_z) + \partial_x\left(\Gamma^2\rho c^2\beta_z\beta_x\right) &=& \rho_{\rm CR}\beta_xB_y  \\
D_tB_y + \partial_x (B_y\beta_x) &=& 0  \\
D_t B_z + \partial_x(B_z\beta_x) &=& 0 \\
D_t(\Gamma^2\rho c^2) + \partial_x(\Gamma^2\rho c^2\beta_x) &=& 0
\end{eqnarray}    
where we neglected all terms proportional either to $\beta_a^2$ or $\beta_s^2$.
\subsubsection{Magneto-acoustic wave propagation}
We consider in this study magneto-acoustic waves displaying magnetic perturbations along the $y-$direction. The induction equation regarding $B_z$ will then be trivial as $B_z$ remains null during the propagation of the wave throughout the precursor.\\
The dominant force in the precursor arises from the presence of CR within the precursor leading to a non neutral plasma. The CR charge density at the shock front is provided by Eq.(\ref{Eq:Profile})
\be
\rho_{\rm CR}(x=0) = \frac{\Gamma^2_{\rm sh}\rho c^2k_*(x=0)}{B_y}
\ee
 Defining $D_x=\Gamma(D_t+\beta_x\partial_x)$ as a Lagrangian derivative, we easily get from the previous set of equations that
\begin{eqnarray}
 D_x\Gamma &=& 0 \nonumber\\
 \Gamma\rho c^2D_x\beta_x &=&  -\rho_{\rm CR}\beta_zB_y  \nonumber\\
 \Gamma\rho c^2D_x\beta_z &=&  \rho_{\rm CR}\beta_xB_y  \\
 \frac{D_x B_y}{B_y} &=& -\Gamma\partial_x \beta_x \nonumber\\
 \frac{D_x \rho}{\rho} &=& -\Gamma\partial_x \beta_x \nonumber
\end{eqnarray}
Using a linear approach, the previous set of equations enables us to get the behavior of small perturbations as for instance magneto-acoustic waves propagating in the precursor. We have 
\begin{eqnarray}
 D_x\delta\Gamma &=& 0 \nonumber\\
 D_x\delta\beta_x &=&  -\frac{\Gamma k_*(\beta_z\delta B_y+B_y\delta\beta_z)}{B_y}  \nonumber\\
 D_x\delta\beta_z &\simeq&  \frac{\Gamma k_*\beta_x\delta B_y}{B_y}  \\
 D_x \frac{\delta B_y}{B_y} &=& -\Gamma\partial_x \delta\beta_x \nonumber\\
 D_x\frac{\delta\rho}{\rho} &=& -\Gamma\partial_x \delta\beta_x \nonumber
\end{eqnarray}
The perturbed momentum equation along the normal to the shock involves two terms in its rhs. Retaining only the term proportional to $\delta\beta_z$ (thus neglecting the transverse velocity) leads to the instability mode found by PLM09 as combining the different equations provides
\be
D_x^3\frac{\delta B_y}{B_y} = \Gamma^3k_*^2\beta_x\partial_x\frac{\delta B_y}{B_y}
\ee
In order to assess the linear growth rate of the magnetic perturbation, we use a standard harmonic description as $D_t\equiv -ik_x\beta_x$ while $\partial_x\equiv ik_x(1-\epsilon)$ where $\epsilon$ is a complex number whose imaginary part leads to the spatial growth rate of the magnetic field. The resulting growth rate of the instability matches the one given by PLM09, namely
\be
 \gamma_x = Im(k_x\epsilon) = \frac{\sqrt{3}}{2}\left(\frac{k_*}{\beta_{\rm sh}}\right)^{2/3} k_x^{1/3} 
 \label{Eq:growthrate}
 \ee
As mentioned earlier, the transverse velocity may be significant as values of $\xi_{\rm CR}$ may reach up to $10\%$ in the precursor. It seems then more realistic that $|\beta_z\delta B_y|\gg |B_y\delta\beta_z|$ regarding the propagation of magneto-acoustic waves. This then leads to a different equation of the evolution of the magnetic perturbation, namely
\be
D_x^2\frac{\delta B_y}{B_y} = -\Gamma\partial_x D_x\delta\beta_x  \simeq \Gamma^2 k_*\beta_z\partial_x\frac{\delta B_y}{B_y}
\ee 
We clearly see here the source of the non-resonant instability which relies on the growth of the velocity perturbation induced by the presence of an initial magnetic perturbation. The magnetic perturbation eventually starts to grow thanks to magnetic induction  which results into an even larger growth of the velocity perturbation and so on.  The growth rate of the magnetic field will then be given by $\epsilon$
\be
\epsilon^2 = -i\frac{k_*\beta_z}{\beta_x^2k_x}(1-\epsilon)
\ee
 which leads in the case of small $\epsilon$ to  
 \be
 \gamma_x = Im(k_x\epsilon) \simeq \left(\frac{k_*\chi k_x}{2\beta_{\rm sh}}\right)^{1/2} 
 \label{Eq:growthrate}
 \ee
 This growth rate is maximal for ultra-relativistic shocks (while being independent of $\Gamma_{\rm sh}$) and decreased for mildly relativistic and non-relativistic shocks (proportional to $\beta_{\rm sh}^{3/2}$ in this regime).  The magneto sonic waves are hence destabilized like in the configuration investigated by \cite{Drury86}, but due to the electromotive force induced by the CR charge. \\
 
 \subsubsection{Alfv\'en waves propagation}
 We have so far only considered magneto-acoustic waves and discarded Alfv\'en waves since they are not prone to this instability. Indeed, in a more general framework we can consider Alfv\'en waves having wave vector $\bm{k}=k_x\bm{e}_x+k_y\bm{e}_y$. The resulting wave is incompressible as only $\delta \beta_z$ and $\delta B_z$ are non-vanishing contributions. The set of RMHD equations then reads
 \begin{eqnarray}
 &&\Gamma^2\rho c^2D_l\delta\beta_x = -\rho_{\rm CR}B_y\delta\beta_z \nonumber \\
 &&\Gamma^2\rho c^2D_l\delta\beta_z = \rho_{\rm CR}B_y\delta\beta_x\nonumber \\
 &&\Gamma^2\rho c^2D_l\delta\beta_y = -\rho_{\rm CR}\beta_x\delta B_z\nonumber \\
 &&D_l\delta B_x =B_y\partial_y\delta\beta_x\nonumber\\
 &&D_l\delta B_y =-B_y\partial_x\delta\beta_x\nonumber\\
 &&D_l\delta B_z =B_y\partial_y\delta\beta_z
 \end{eqnarray} 
 where $D_l=\Gamma(D_t+\beta_x\partial_x+\beta_y\partial_y)$ stands for the 2D Lagrangian derivative. We retain only terms from the perturbations $\delta\beta_z$ and $\delta B_z$ plus the term involved in the evolution of $\delta\beta_z$, namely $\delta\beta_x$. Combining the first two equations leads to the evolution equation of $\delta\beta_z$ 
 \be
 D_l^2\delta\beta_z = -\frac{\rho_{\rm CR}^2B_y^2}{\Gamma^2\rho^2c^4}\delta\beta_z
 \ee
 which does not have any exponentially diverging solutions. Thus, Alfv\'en waves propagating through the precursor are found to be stable (see also PLM09).
\section{RMHD waves propagating in the precursor of a relativistic shock}
\label{RMHD}
 The presence of the CR electrical charge density (and the associated current) is the key origin of the non-resonant instability as it is able to destabilize MHD waves. We will first present the RMHD framework and the numerical tool we used to perform the non-linear simulations described in this work. In a second step, we present a linear description of the instability in a restrained but nevertheless interesting case, namely a transverse magnetosonic wave flowing towards the shock front. 

\subsection{Numerical code description and initial setup}
 The MPI-parallelized, Adaptive Mesh Refinement Versatile Advection Code (\textsc{MPI-AMRVAC}) is a multi-dimensional numerical tool devoted to solve conservative equations (\citet{vdHol08,keppens12}) using finite volume techniques and a dynamically refined grid. The \textsc{MPI-AMRVAC} package includes sets of equations  as hydrodynamical or magnetohydrodynamical equations either in a classical or relativistic framework. For the simulations displayed in this paper, we used a second-order Harten-Lax-van Leer Contact (HLLC) solver linked to a Koren slope limiter \citep{Kore93} to make sure we employ a robust but accurate scheme. \\
 The AMR refinement strategy can be controlled by several means within the \textsc{MPI-AMRVAC} framework,  such as by Richardson extrapolation to future solutions or using instantaneous quantifications of the normalized second derivatives, or by a user controlled criterion or actually both (\citet{keppens12}). For the purpose of our simulations, we simply choose to enforce the maximal refinement to follow the wave propagating throughout the precursor of the shock knowing its velocity as given by Eq.(\ref{Eq:propagation}). 
 The base level is filled with blocks of equal size which can be divided into $2^D$ child grids having the same amount of grid cells than the parent grid ($D$ being the dimension of the grid). The structure of the grid will then be similar to an octree for three dimensional calculations.  In the context of a perpendicular relativistic shock, the aforementioned  set of RMHD equations simplifies as ${\bf B}\cdot\bm{\beta}=0$. Assuming the magnetic field to be oriented along the $y$ direction (parallel to the shock front), we obtain for 1D simulations the following set of equations, namely
\begin{eqnarray}
\partial_tD + \partial_x (D\beta_x) &=& 0  \\
\partial_tS_x + \partial_x\left(S_x\beta_x + P_{\rm tot}\right) &=& -\rho_{\rm CR}\beta_zB_y  \\
\partial_tS_z + \partial_x\left(S_z\beta_x\right) &=& \rho_{\rm CR}\beta_xB_y  \\
\partial_tB_y + \partial_x (B_y\beta_x) &=& 0  
\end{eqnarray}
where we took into account the CR charge effect upon the momentum of matter.   
Considering the physical conditions prevailing nearby relativistic shocks, PLM09 assumed that both the magnetic pressure and thermal pressure can be neglected compared to mass energy density. From a numerical point of view this is a quite difficult feature to deal with as the energy equation is likely to provide numerical errors leading to negative thermal pressure. In order to avoid that problem, we decided to  set a simple isothermal relation linking thermal pressure to mass density, $P=\rho c_S^2$ where $c_S\ll c$ is the sonic speed in the matter rest frame. The procedure to switch from conservative to primitive variables ($\rho, \bm{\beta}, {\bf B}, P)$ has then to be revisited compared to the one presented in \citet{vdHol08} (see Appendix A). In one dimensional computations dealing with shocks having magnetic field parallel to the shock front,  the conservation of magnetic flux is naturally enforced as $\partial_xB_x=0$.\\
We set up the initial conditions by following the equilibrium presented in Sect.(\ref{Loreconst}). The equilibrium is entirely controlled by two free parameters, namely the shock Lorentz factor $\Gamma_{\rm sh}$ and the amount of kinetic energy of the shock transferred into the CR population $\xi_{\rm CR}$. The parameter $\beta_a$ and $\beta_s$ are also free but their actual value do not influence the instability behavior provided it remains very small compared to unity (we specifiate their value for each simulation in the following). One final ingredient is required to close the initial setup, namely the shape of the cosmic ray charge density. In order to simplify the instability study, we choose $\rho_{\rm CR}$ to be constant within the precursor and vanishing outside. We then choose to set 
\begin{equation}
\rho_{\rm CR}(x) = 
\begin{cases}
\displaystyle\frac{\xi_{\rm CR}\Gamma_{\rm sh}(\Gamma_{\rm sh}-1)\rho_oc^2}{\ell_{\rm CR}B_o}\tanh^3\left(\frac{x_*-x}{\ell_o}\right), &  0 < x \leq x_* \\
0 \ , &  x \geq x_*
\end{cases}
\end{equation}
where $l_o$ is a constant such that $\ell_o=0.03\ell_{\rm CR}$ and $B_o$ and $\rho_o$ are the magnetic field and mass density at the location of the shock $(x=0)$. The computational domain ranges from $x=0$ where the shock front is located, to $x=1.4\ell_{\rm CR}$. Let us note that we set $x_*=1.1\ell_{\rm CR}$ to be the border between the precursor of the shock and the unaffected ISM medium. The vertical velocity $\beta_z$ is linked to the charge density profile as stated by eq.(\ref{Eq:vertequi}) which provides
\begin{eqnarray}
u_z(x)&=& -\frac{(\Gamma_{\rm sh}-1)\xi_{\rm CR}\ell_o}{\ell_{\rm CR}}\left(\frac{1}{2}\text{sech}^{2}\left(\frac{x-x_*}{\ell_o}\right)\right. \nonumber\\
&&+\left.\ln\left\{\cosh\left(\frac{x-x_*}{\ell_o}\right)\right\}-\frac{1}{2}\right)
\end{eqnarray}
The velocity along the normal to the shock can be derived from the aforementioned equilibrium, namely when $\Gamma=\Gamma_{\rm sh}$. The previous equilibrium is consistent with $u_y=0$ so $u_x(x)=-\sqrt{\Gamma_{\rm sh}^2-1-u_z(x)^2}$. We also showed that the magnetic field components $B_x$ and $B_z$ are null. The remaining quantities ($B_y$ and $\rho$) are easily constrained by the mass conservation and induction equation leading to
\begin{eqnarray}
u_x(x)B_y(x)&=&-(\Gamma^2_{\rm sh}-1)^{1/2}B_{\rm ISM}\\
u_x(x)\rho(x)&=&-(\Gamma^2_{\rm sh}-1)^{1/2}\rho_{\rm ISM}
\end{eqnarray}  
where $B_{\rm ISM}$ is the value of the magnetic field in the unperturbed region ahead of the precursor. We choose to set the density of this region $\rho_{\rm ISM}=1$ so that $B_{\rm ISM} = \beta_a\sqrt{4\pi c^2\Gamma^2_{\rm sh}}$ in our simulations.  As an instance, we displayed in Fig.(\ref{F1}) the variation of the various physical quantities for an ultra-relativistc shock having $\Gamma_{\rm sh}=100$, $\xi_{\rm CR}=10\%$ and $\beta_a^2=10^{-10}$ and $\beta^2_s=10^{-8}$.\\
The analytical equilibrium derived in Sect(\ref{Loreconst}) is only an approximation of the actual equilibrium since we discarded the magnetic and thermal pressure terms involved in the momentum equation (Eq.\ref{Eq:Motion}). The neglected pressure terms are very small compared  to the force associated with the electromotive field 
\be
\left|\frac{B_y\partial_xB_y}{4\pi\rho_{\rm CR}\beta_zB_y}\right| \simeq \frac{\beta_a^2}{\beta^2_{\rm sh}} \ll 1 \ ; \left|\frac{\partial_xP}{\rho_{\rm CR}\beta_zB_y}\right| \simeq \frac{\beta_s^2}{\beta^2_{\rm sh}} \ll 1
\ee
In the context of the numerical equilibrium presented in fig.(\ref{F1}), the previous ratio are equal to $10^{-10}$ and $10^{-8}$ respectively. From a numerical point of view this analytical equilibrium is not suitable for our simulations as the magneto-acoustic wave entering the precursor region displayed velocity disturbances smaller than the previous ratios. We then have to implement a numerical approach that enforces an exact equilibrium to machine precision. To do so we employ  the same technique than \citet{Mehe10}, namely we compute the electromotive source terms  in a two step fashion. During the first step, we calculate the numerical errors produced by the non accurate initial conditions presented in Sect.(\ref{Loreconst}) such that
\begin{eqnarray}
D_t\bm{S}_{\rm i} &+& {\bf \nabla}\cdot\left(\bm{S}_{\rm i}\bm{\beta}_{\rm i} - \left(\frac{\bm{B}_{\rm i}}{\Gamma^2_{\rm i}}+(\bm{\beta}_{\rm i}\cdot\bm{B}_{\rm i})\bm{\beta}_{\rm i}\right)\frac{\bm{B}_{\rm i}}{4\pi}\right. \nonumber \\ 
&+&\left. P_{\rm tot,i}\bf{I}\right) -\rho_{\rm CR}(\bm{\beta}_{\rm i}\times\bm{B}_{\rm i})= \bm{\Phi}_{i}
\end{eqnarray}
where the subscript $i$ stands for initial conditions quantities. Once the numerical errors are calculated, we used them into the second step dealing with the actual physical conditions, namely 
 \begin{eqnarray}
D_t\bm{S} &+& {\bf \nabla}\cdot\left(\bm{S}\bm{\beta} - \left(\frac{\bm{B}}{\Gamma^2}+(\bm{\beta}\cdot\bm{B})\bm{\beta}\right)\frac{\bm{B}}{4\pi}\right. \nonumber \\ 
&+&\left. P_{\rm tot}\bf{I}\right) = \rho_{\rm CR}(\bm{\beta}\times\bm{B})+ \bm{\Phi}_{i}
\end{eqnarray}
Let us note that this procedure has to be repeated each time the adaptive mesh evolves since any modification in the grid hierarchy would lead to  different values of $\bm{\Phi}_{\rm i}$. \\ 
At the beginning of the simulation we apply the aforementioned procedure during the first time step (and at each following time step where the grid hierarchy chages) and at the beginning of the second time step we add to the equilibrium physical quantities the perturbation associated to a magneto-acoustic wave packet  in the region ahead of the upstream medium where no CR charge is present. More specifically we add perturbation to the mass density, magnetic field and velocity field. The spatial extent of the wave packet is equal to $24\pi/k_x$ (twelve wavelengths) as can be seen on the upper left plot of Fig.(\ref{F2}). The perturbation is coherent with a magneto-acoustic wave whose wave vector is ${\bf k}=k_x{\bf e}_x$ in the shock rest frame.  The detailed expression of the implemented perturbation is provided by Eq.(\ref{Initperturb}).  The boundary conditions are simply continuous boundary where all physical quantities are copied in the ghost cells.


\subsection{A reference simulation of the instability}
\label{Sect:Figu}
\subsubsection{The basic stages of the instability}
\begin{figure*}
\centering
\[ \begin{array}{cc} 
 \includegraphics[width=0.5\textwidth]{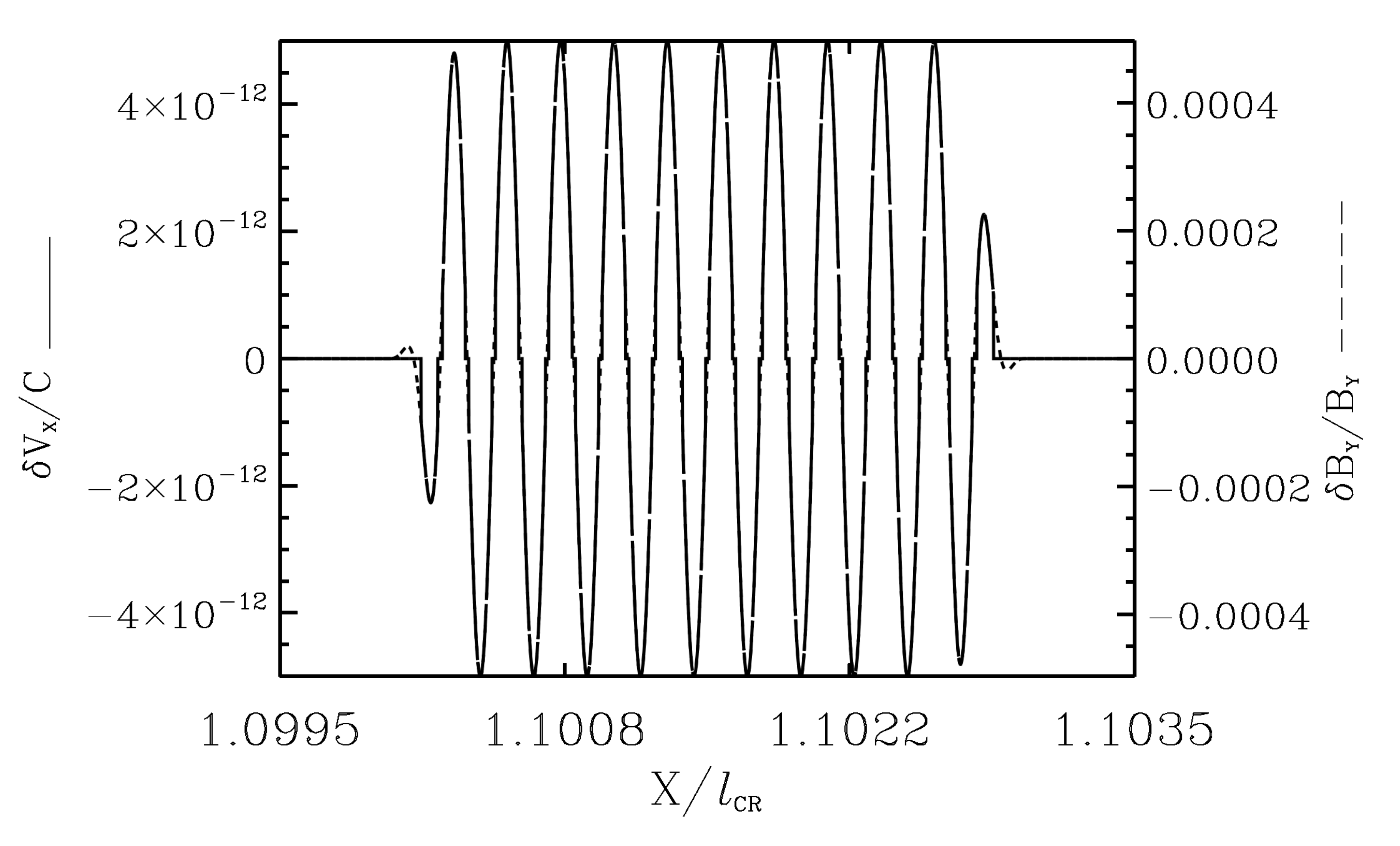} &
 \includegraphics[width=0.5\textwidth]{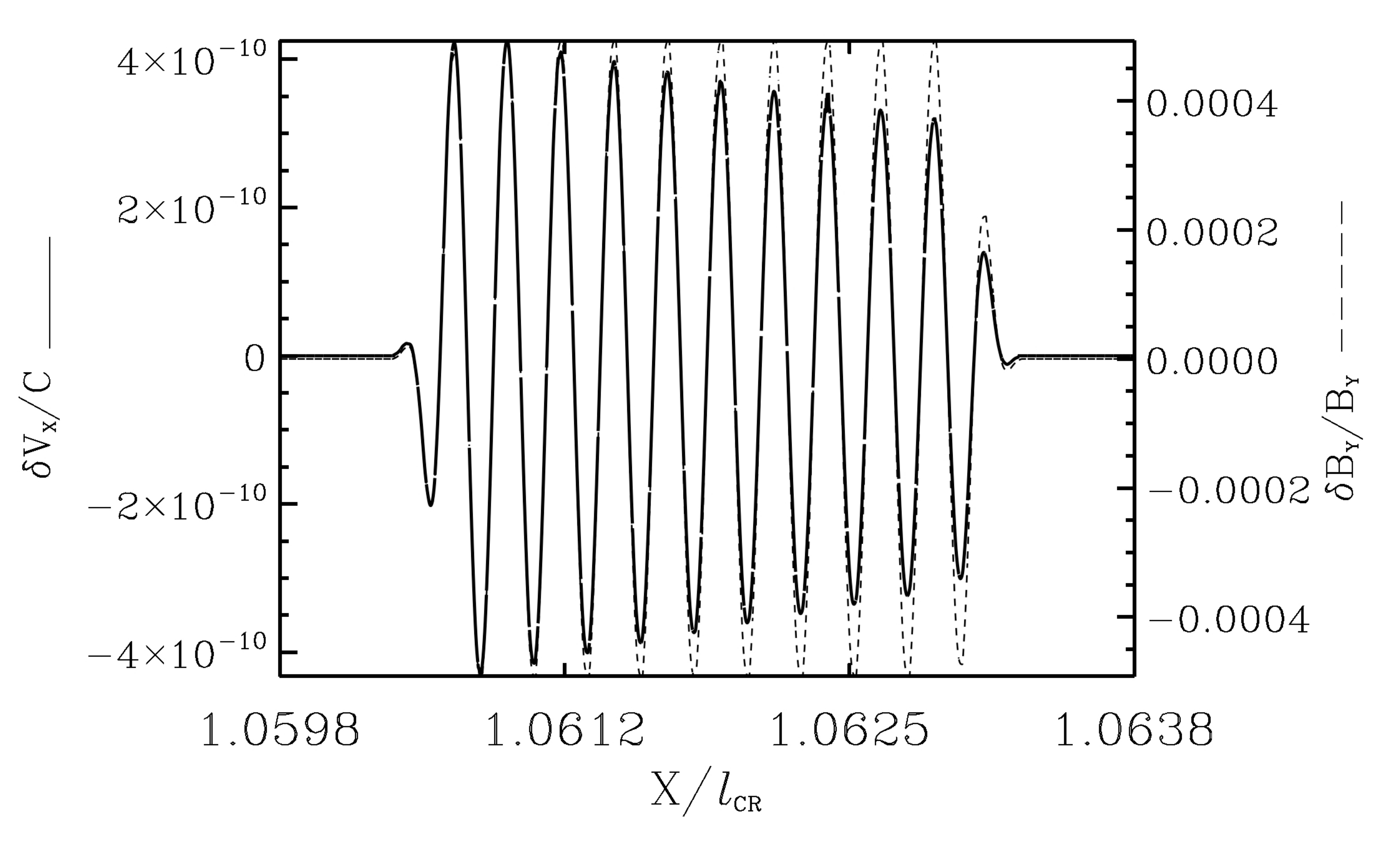} \\
  \includegraphics[width=0.5\textwidth]{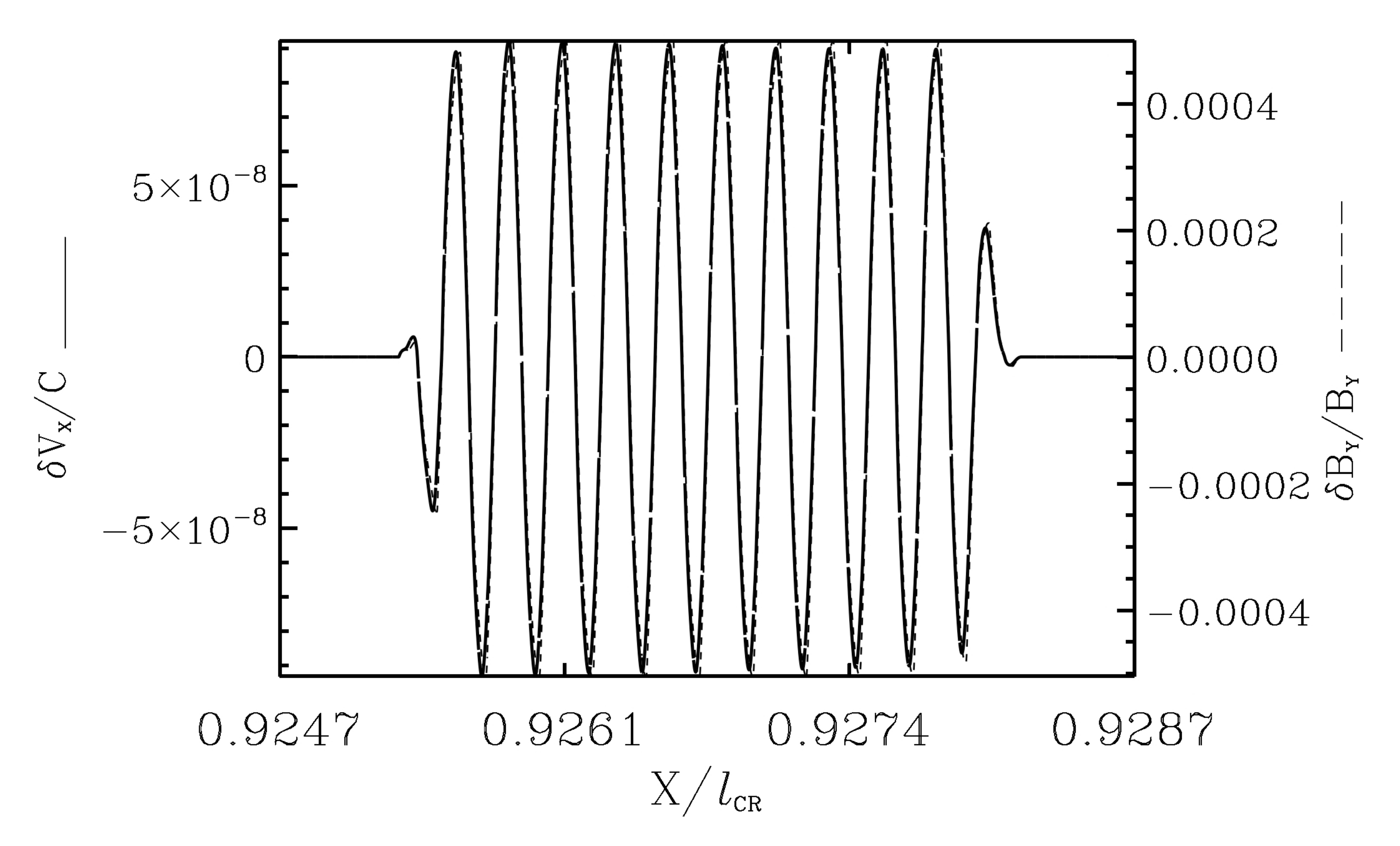} &
 \includegraphics[width=0.5\textwidth]{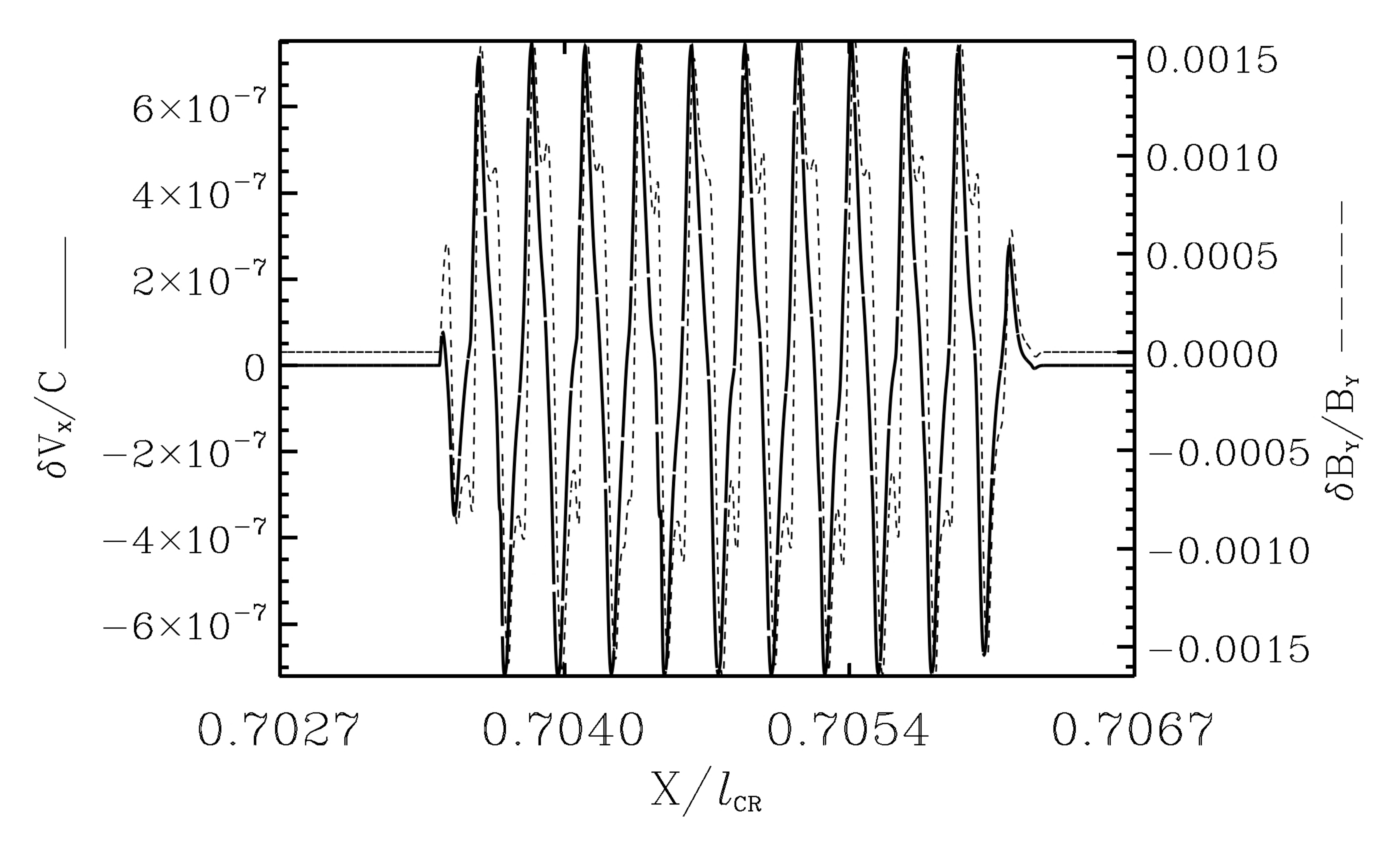} \\
  \includegraphics[width=0.5\textwidth]{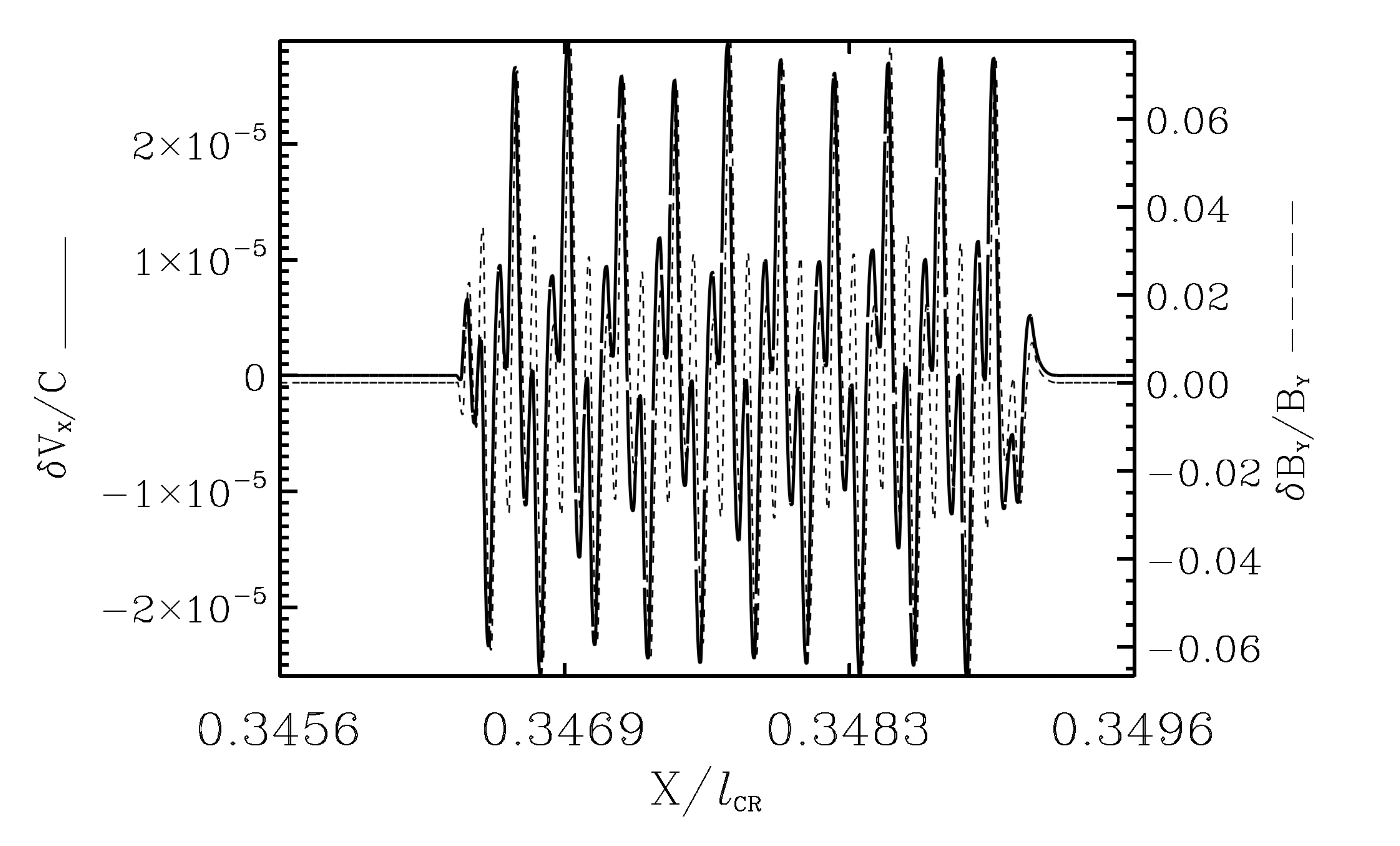} &
 \includegraphics[width=0.5\textwidth]{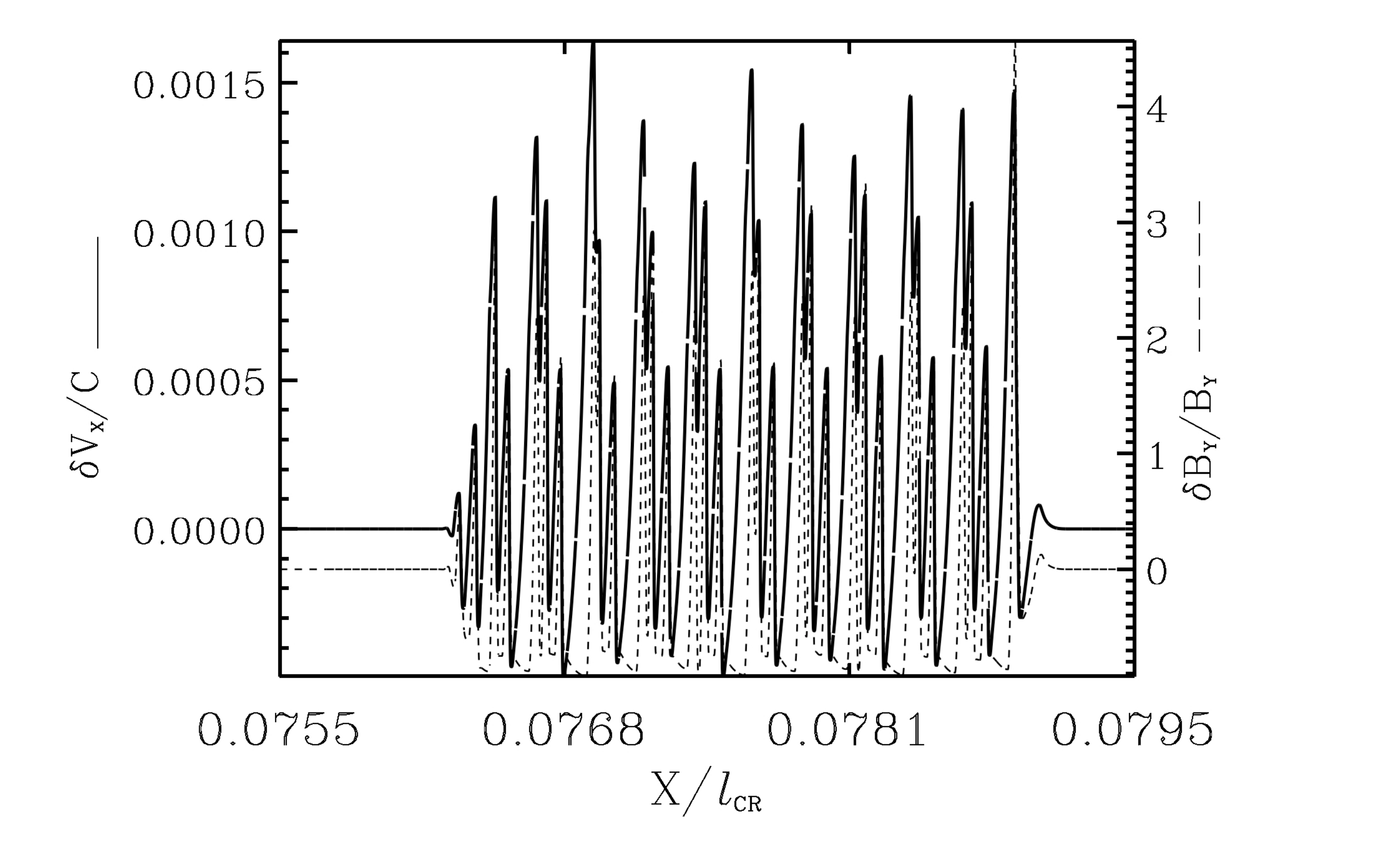} 
  \end{array}  \]
 \caption{Snapshots of the reference simulation presented in Sect.(\ref{Fiducial}). We have displayed the parallel velocity perturbation together with the magnetic perturbation at locations related to the various stage of the instability. At first (top left panel) the initial magneto-acoustic perturbation is entering the cosmic ray dominated region then (top right) the velocity perturbation is increasing while the magnetic perturbation remains unchanged. The third step (middle left) shows the beginning of the magnetic field amplification when both perturbations start to be out of phase. The fourth stage (middle right) shows that the magnetic amplification starts to modify the shape of the magnetic perturbation because of the  phase shift (being different from $\pi/2$). The fifth stage (bottom left) shows that once the magnetic perturbation is modified, it then affects the shape of the velocity perturbation.  The final stage corresponds to a fully non-linear stage where the magnetic field is completely dominated by the magnetic disturbance. The figures also display a time sequence evolving from the upper left (perturbation entering the precursor at $X_{\rm max}/\ell_{\rm CR}=1.1$) to lower right (perturbation reaching the shock front at $X=0$) with timescales given by $(X_{max}-X)/\beta_{sh}$, where $X_{max}/\ell_{\rm CR}=1.1$.}
\label{F2}
\end{figure*}
The initial perturbation entering the CR charge dominated region is prone to the additional force arising from the CR charge density. As it can be seen from Eq.(\ref{Initperturb}), the initial perturbed velocity $\delta\beta_x$ is very small compared to $\delta B_y/B_y$ so the growth of the velocity perturbation will not induce any amplification of the magnetic disturbance right from the start. During this preliminary stage, it is possible to infer the  growth of the velocity perturbation by looking at the linearized SRMHD equations. Taking into account the mass density conservation and the induction equations leads to
\begin{eqnarray}
\partial_t\delta\rho &\simeq &-\beta_x\partial_x\delta\rho -\partial_x\delta\beta_x\\
\partial_t\delta B_y  &\simeq &-\beta_x\partial_x\delta B_y -\partial_x\delta\beta_x
\end{eqnarray}
which are fully consistent with a pure advection motion as long as $\delta\beta_x$ and $\delta B_y/B_y$ remain synchronized. On the other hand, the linearized equation of motion lead to the following relations
\begin{eqnarray}
\partial_t \delta\beta_x + \beta_x\partial \delta\beta_x +\frac{\beta_F}{\Gamma_{\rm sh}^2}\partial_x \frac{\delta B_y}{B_y} &\simeq & -\xi_{\rm CR}\frac{\beta_z\delta B_y}{\ell_{\rm CR}B_y} \\
\partial_t \delta\beta_z + \beta_x\partial \delta\beta_z +\frac{\beta_F}{\Gamma_{\rm sh}^2}\partial_x \frac{\delta B_y}{B_y} &\simeq& \xi_{\rm CR}\frac{\beta_x\delta B_y}{\ell_{\rm CR}B_y} 
\end{eqnarray}
 The presence of the cosmic ray charge density alters the nature of these equations since they are no longer consistent with a pure advection motion. The amplitude of the velocity disturbances increases while the wave is propagating and we can see that the velocity perturbation is then
 (assuming the disturbances propagates at speed very close to $\beta_x$ which remains almost constant since $\beta_z$ is very small at the outer edge of the CR charge dominated region)
\begin{eqnarray}
 \delta \beta_x (x) &\simeq& \delta \beta_x(x_*) + \frac{\beta_z^2(x)}{2\beta_{\rm sh}}\frac{\delta B_y}{B_y}(1+\beta^2_x) \nonumber\\
 \delta \beta_z(x) &\simeq&\beta_z(x)\frac{\delta B_y}{B_y}\left(1-\frac{\beta_z^2(x)}{2}\right)
 \label{Eq:assess}
 \end{eqnarray}
 where we used the fact that $\rho_{\rm CR} \simeq (\Gamma_{\rm sh}\rho c^2\partial_xu_z)/B_y$ (see Eq.\ref{Eq:coldequi}). 
 During this first stage, only the velocity disturbances are amplified while the magnetic perturbation remains unchanged while propagating. Nevertheless the growth of the velocity disturbance has an impact on the advection velocity of the magnetic perturbation as can be seen from the linearized induction equation, namely
 \begin{eqnarray}
 \partial_t\delta B_y + \left(\beta_x +\frac{k_x}{|k_x|}\frac{\beta_F}{\Gamma_{\rm sh}^2} + \frac{\beta_z^2(x)}{2}(1+\beta^2_x)\right)\partial_x \delta B_y= 0
  \end{eqnarray}
 \begin{figure}
\centering
 \includegraphics[width=0.5\textwidth]{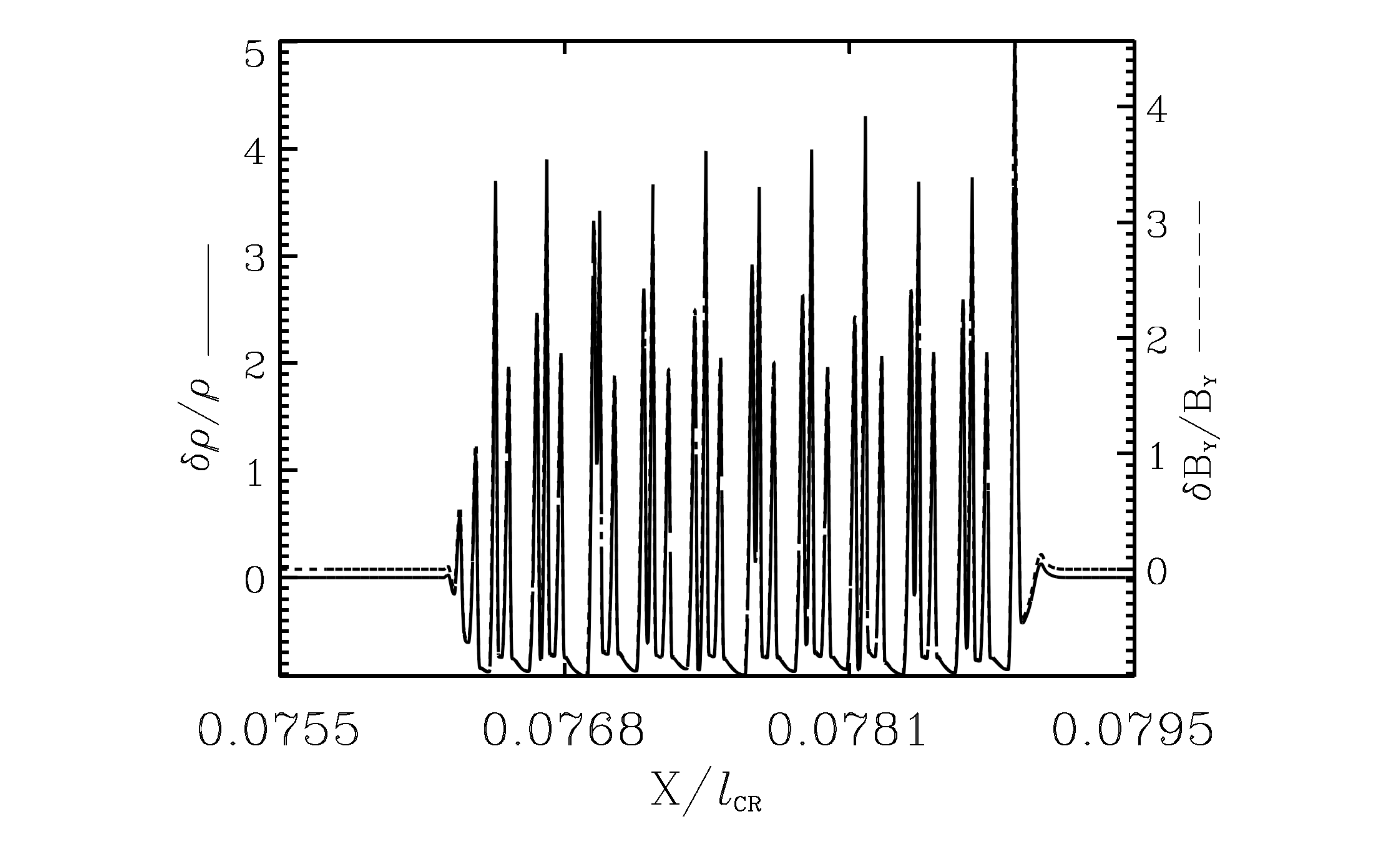} 
  \caption{The mass density and magnetic perturbations at the final stage of the simulation. The system has encountered the saturation limit expected from the amplification of a magneto-acoustic wave, namely the regime where $\delta\rho/\rho\rightarrow -1$. We can see that the initial wave has been amplified to a point where the density of the medium exhibits very sharp variations that may be associated with shock fronts within the precursor.} 
\label{F3}
\end{figure}

 The advection velocity of the magnetic disturbance is clearly depending on the transverse velocity of the flow induced by the CR charge density. Since the advection speed of the velocity disturbances remains constant, the magnetic and mass density perturbations begin to be out of phase with  the velocity disturbances which leads to an amplification of both magnetic and mass density perturbations. The wave is then entering a second stage where the instability arises from both magnetic and velocity exponential growth. In order to assess the location of the transition between the two stages, we may look at the previous equation where falling back to the general expression, we have
 \be
 \partial_t\delta B_y + \beta_x\partial_x\delta B_y + B_y\partial_x\delta \beta_x \sim \frac{d\delta B_y}{dt} + B_y\partial_x\delta \beta_x = 0
 \ee 
 The transition between the pure advection stage and the magnetic field amplification stage occurs when the velocity perturbation is sufficiently large to generate a variation of the magnetic perturbation of the order of $\delta B_y/B_y$ when crossing the entire CR charge dominated region, namely when $\Delta \delta B_y/B_y \sim \delta B_y/B_y$. This leads to a velocity perturbation threshold 
 \be
 \delta\beta_x^{\rm Trans} \sim \frac{\beta_{\rm sh}\lambda}{\ell_{\rm CR}}\frac{\delta B_y}{B_y} \ {\rm and} \ \beta_z^{\rm Trans} \sim \sqrt{\frac{\beta_{\rm sh}\lambda}{\ell_{\rm CR}}}
 \label{Eq:magamp}
 \ee
 where $\lambda$ is the typical wavelength of the magneto-acoustic wave. It is obvious that shorter wavelength waves will ignite 
 the magnetic field amplification earlier as the threshold relation is met earlier in the wave propagation.\\
 At a certain point, a third stage is expected as the phase shift between magnetic and velocity disturbances continues to increase during the second stage. Once the offset is sufficiently large, the magnetic perturbation will modify the shape of the velocity perturbation since its maximal growth will  occur in parts of the wave where the velocity disturbance is not maximal. The modification of the velocity perturbation will then alter the magnetic disturbance profile and so on. This third stage is consistent with a fully non-linear dynamics.

 \subsubsection{A reference numerical simulation in an ultra-relativistic shock}
 \label{Fiducial}
 In order to illustrate the aforementioned stages of the instability, we designed a reference simulation matching all physical conditions assumed in the linear analysis, namely where $B^2/4\pi\sim 10^2 P \sim 10^{-8}\rho c^2$. We set up  an ultra-relativistic shock with $\Gamma_{\rm sh}= 100$, ensuring that the initial magneto-acoustic perturbation is such that $\delta\beta_x\ll \delta B_y/B_y$. In order to get a large growth rate, we set the ratio of CR energy to shock energy $\xi_{\rm CR} = 10\%$ \citep{Milos06}. The various physical quantities are initially set according to the equilibrium presented in Sect.(\ref{Loreconst}) and displayed in Fig.(\ref{F1}). 
 \begin{figure}
\centering
 \includegraphics[width=0.5\textwidth]{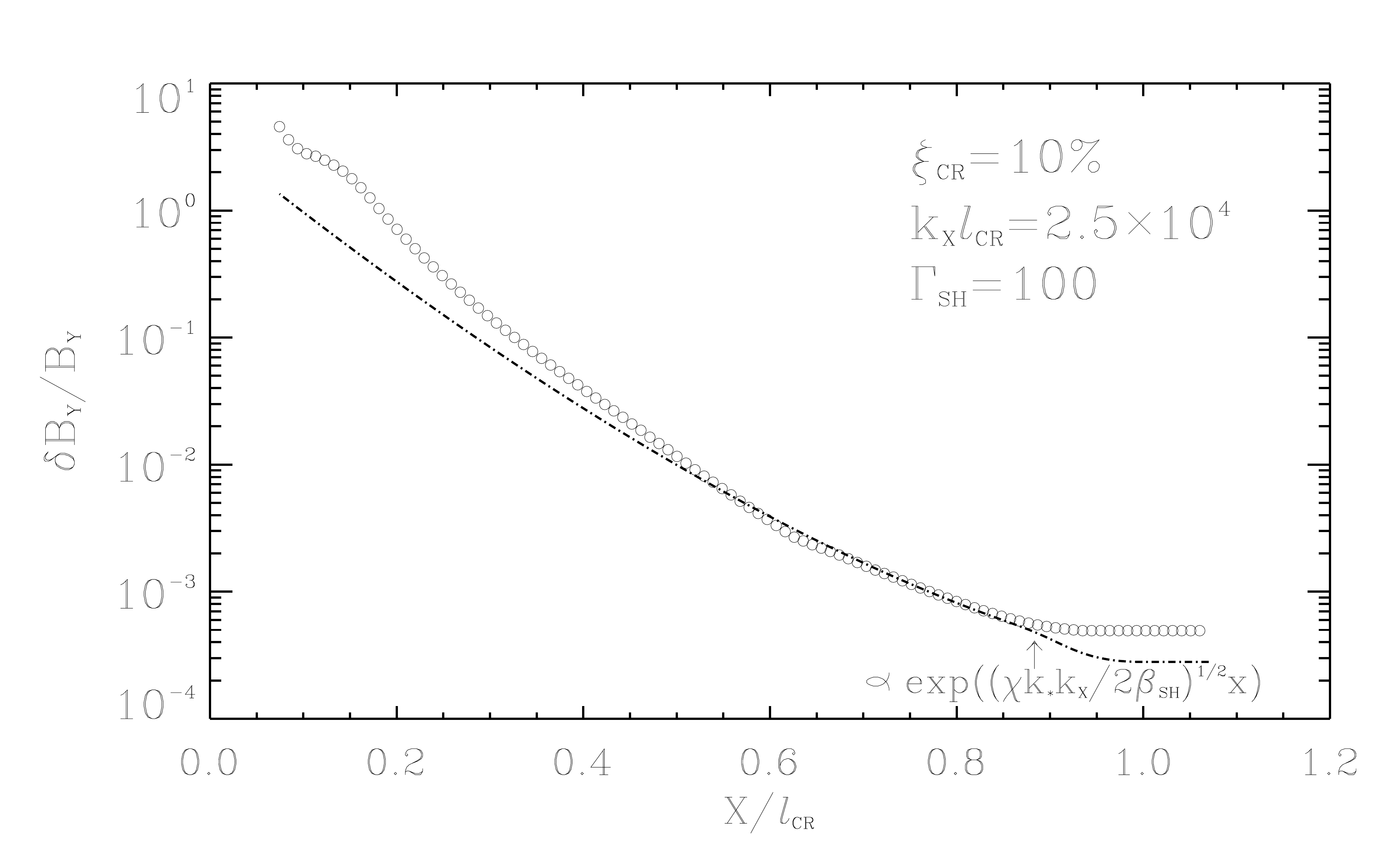} 
  \caption{The maximum amplitude of the magnetic perturbation as a function of the location of the perturbation. The exponential growth of the perturbation matches the linear estimate of the instability. } 
\label{F4}
\end{figure}
 The numerical equilibrium is achieved within machine precision according to the procedure presented in the previous section. 
 We start the simulation with the previous setting and add a magneto-acoustic wave according to Eq.(\ref{Initperturb}) at the outer right hand side border of the upstream region, namely where $\rho_{\rm CR}=0$. The wave vector of the perturbation is such that $k_x\ell_{\rm CR}\simeq 2.5\times 10^4$. In order to capture the dynamics of the instability, we used the adaptive mesh refinement technique. The base level of the simulations has $1700$ cells while $10$ sub-grid levels were used, reaching an effective resolution of near $8\times 10^{-7}\ell_{\rm CR}$. Computing the crossing  from right to left of the entire region by the wave is quite computationally challenging since it requires several millions iterations.\\

 \begin{figure*}
\centering
 \includegraphics[width=\textwidth]{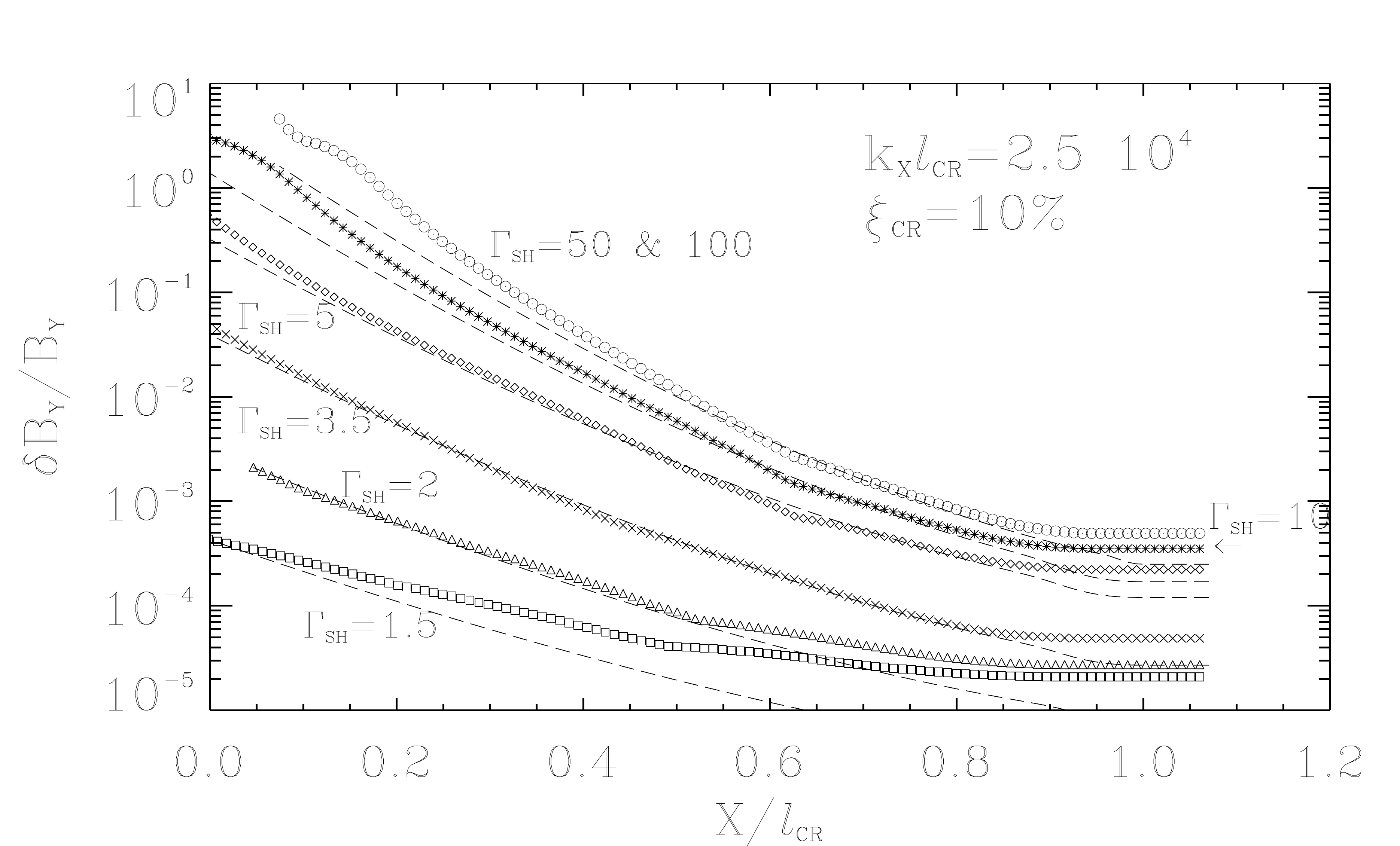} 
  \caption{Same plot than in Fig.(\ref{F4}) but for various Lorentz factors ranging from $\Gamma_{\rm sh}=100$ down to $\Gamma_{\rm sh}=1.5$. The spatial growth rates corresponds to similar simulations than the one presented in Sect.(\ref{Fiducial}) but for various shock velocity ($\Gamma_{\rm sh}=100, 50, 10, 5, 3.5, 2, 1.5$). The linear growth rate provided by Eq.(\ref{Eq:growthrate}) is displayed using dashed lines. We clearly see that once entering the non-relativistic regime, the growth rate of the magneto-acoustic waves tends to zero thus no instability is observed.  The numerical curves differ from the linear growth ratesat the beginning of the propagation since the instability is not at work right from the start of the simulations but instead has to get through the various stages described by Sect.(\ref{Sect:Figu}). We then adjust the linear curves to fit the numerical calculations by tuning the linear original amplitude of the instability.} 
\label{F5}
\end{figure*}
 We displayed in Fig.(\ref{F2}) various snapshots of both the velocity and magnetic perturbations from the numerical calculation. Any perturbed quantity $A$  is computed as $\delta A= A(t)-A(t=0)$ where $A(t=0)$ is the quantity as defined by the initial equilibrium.
 In the top left panel 
 we displayed the initial perturbation setup where we can see the ratio of the relative amplitudes is $10^8=\Gamma_{\rm sh}^2/\beta_F$. On the top right panel, the wave has entered the charge layer but is located in the region where $\rho_{\rm CR}(x)$ is increasing.  The magnetic perturbation remains unchanged while the velocity perturbation has grown already. The amplitude of the velocity perturbation matches the assessment from Eq.(\ref{Eq:assess}) very well. The middle left plot shows the wave at a stage where the magnetic perturbation is starting to grow. At this point, the magnetic perturbation is slightly out of phase with the velocity component as expected. Let us note that the magnetic field amplification occurs when the ratio between the velocity perturbation to the magnetic perturbation is of the order of $2\times 10^{-4}$ which is close to the set wavelength of the magneto-acoustic wave, namely $\lambda=2.5\times 10^{-4}\ell_{\rm CR}$. This is in agreement with the magnetic amplification threshold identified in Eq.(\ref{Eq:magamp}). 
 
 The middle right panel shows the two perturbations at a later stage where the phase shift between the two components becomes quite significant. This shift induces a modification of the perturbation profile. The bottom left plot shows the status of the system while entering the non-linear regime, namely when $\delta B_y/B_y$ is no longer $\ll 1$. The perturbation profiles are completely distorted by non-linear effects. Finally the bottom right plot shows the system at its final stage, namely where saturation occurs. In this final snapshot the magnetic perturbation has reached an amplitude larger than the original magnetic field, leading to a modest but significant magnetic amplification (up to $5$ times the initial magnetic field). As already mentioned, mass density and magnetic field are controlled by similar equations so it is expected that the  mass density reaches a similar state than the magnetic field. As shown in Fig.(\ref{F3}), it is verified by the numerical simulation since both mass density and magnetic field perturbations are very similar. The magnetic field fluctuations significantly exceed those of the initial perturbation such that $\delta B/B > 0$, meaning that the magnetic flux has increased and that the magnetic field has been amplified. The amplification of the magnetic field and mass density leads to regions where $\delta\rho/\rho\rightarrow -1$, namely regions whose mass density tends toward zero. In our simulations, this regime ultimately leads to a numerical instability, since true zero densities cannot be handled by the MHD code, and fiducial fixes to avoid negative densities become activated.\\
 
 The temporal evolution of the magnetic perturbation amplitude is displayed in Fig.(\ref{F4}). The spatial growth of the magnetic disturbance does not show a constant exponential growth during the precursor crossing. This was expected as the spatial growth rate inferred from linear analysis involves functions $\chi(x)$ and $k_*(x)$ depending of the distance $x$ from the shock front.  We displayed on Fig.(\ref{F4}) the growth rate predicted from Eq.(\ref{Eq:growthrate}). Both the numerical and theoretical growth rates match quite well during the first part of the simulation while the perturbation remains very small compared to the mean magnetic field. Once the perturbation has grown, the numerical growth rate becomes larger than the linear prediction. This is not surprising since at this stage $\delta B_y/B_y$ has reached a relative amplitude of a few percents and the linear assumption no longer holds. We can observe at the end of the simulation that the saturation process occurs leading to a slight flattening of the growth rate.
 \begin{figure*}
\centering
 \includegraphics[width=\textwidth]{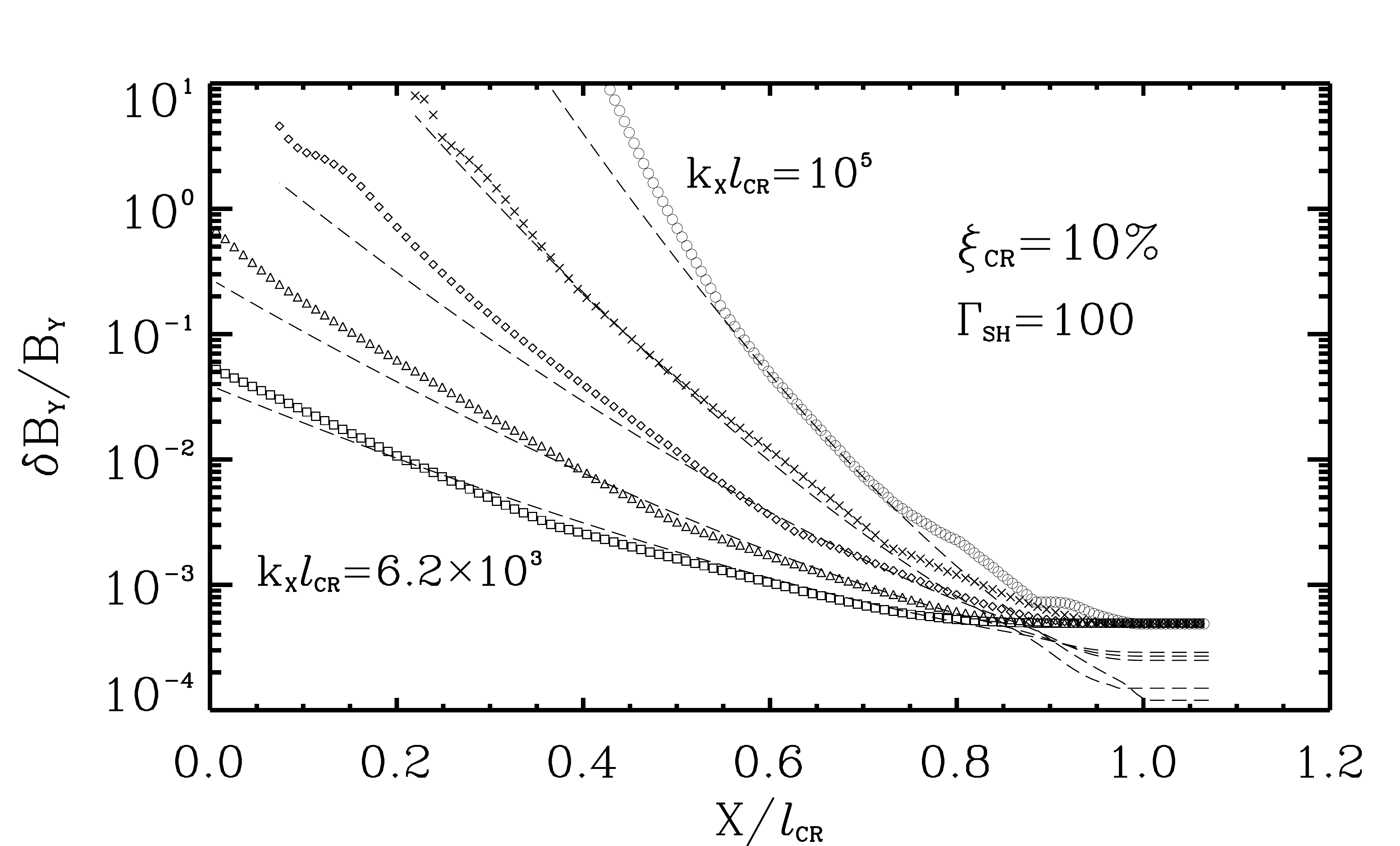} 
  \caption{Same plot than in Fig.(\ref{F4}) but for various wavenumber ranging from $k_x\ell_{\rm CR}=6.25\times 10^3$ up to $k_x\ell_{\rm CR}=10^5$. The spatial growth rates correspond to similar simulations than the one presented in Sect.(\ref{Fiducial}) except for the wavenumber of the perturbation ($k_x\ell_{\rm CR}=6.25\times 10^3, 1.25\times 10^4, 2.5\times 10^4, 5\times 10^4$ and $10^5$). The linear growth rates are obtained from  Eq.(\ref{Eq:growthrate}). The numerical growth rates obtained from simulations follow the predicted linear growth rate which is proportional to $k_x^{1/2}$. } 
\label{F6}
\end{figure*}
 \subsubsection{Non-resonant streaming instability; from ultra-relativistic to non-relativistic shocks}
 
 We have run similar simulations to the one presented in the previous section but with various shock velocity and Lorentz factors. The growth rate predicted by the linear analysis is 
 \be
 \gamma_x(x) = Im(k_x\epsilon) \simeq \left(\frac{k_*(x)\chi(x) k_x}{2\beta_{\rm sh}}\right)^{1/2} 
 \ee 
 where $\chi(x)$ and $k_*(x)$ are functions such that the spatial growth reads
 \be
 \gamma_x(x=0) \simeq \frac{\xi_{\rm CR}(\Gamma_{\rm sh} -1)}{\Gamma_{\rm sh}}\left(\frac{k_x}{2\beta_{\rm sh} l_{\rm CR}}\right)^{1/2}
 \ee 
 In the case of ultra-relativistic shocks, we easily get that the growth rate is independent of the shock velocity as 
 \be
 \gamma_x (x=0)\simeq  \xi_{\rm CR}\left(\frac{k_x}{2\ell_{\rm CR}}\right)^{1/2}
 \ee
 The mildly relativistic shock regime is more cumbersome. Since we can assume $\beta_{\rm sh}$ to remain close to unity in this regime, we see that the growth rate is 
 a slowly decreasing function of $\Gamma_{\rm sh}$.\\ 
 For non-relativistic shocks, the spatial growth rate reads
 \be
 \gamma_x(x=0)\simeq \xi_{\rm CR}\left(\frac{\beta_{\rm sh}}{2}\right)^{3/2}\left(\frac{k_x}{\ell_{\rm CR}}\right)^{1/2}
 \ee
 which is dependent on the shock velocity. According to the previous expression, we thus expect the instability efficiency to decrease as the shock velocity decreases.\\

We have displayed in Fig.(\ref{F5}) the spatial growth rates obtained from various simulations with different Lorentz factors. The top curve (displayed with circles) corresponds to two simulations with $\Gamma_{\rm sh}=100$ and $50$: both curves are quasi identical 
proving that for ultra-relativistic shocks, the spatial growth rate is independent of the shock velocity or Lorentz factor (the linear growth rate relative difference between the two simulaton is less than $0.7\%$ so that we only plotted the linear growth rate related to the $\Gamma_{\rm sh}=100$ simulation).  A similar simulation but with $\Gamma_{\rm sh}=10$ is plotted using stars. It exhibits a similar behavior than the previous curves but with a slightly smaller slope, as expected since this simulation is consistent with a mildly relativistic shock. The same behavior is observed for shocks having $\Gamma_{\rm sh}=5, 3.5$ and $2$. Let us note that all curves match well the linear estimate from Eq.(\ref{Eq:growthrate}) as long that the relative magnetic perturbation remains small compared to unity. The last plot (displayed with squares) stands for a simulation of a shock whose velocity is $\beta_{\rm sh}\simeq 0.74$ ($\Gamma_{\rm sh}=1.5$). The slope of the plot is stalling in that case and almost no wave amplification occurs during the crossing of the precursor. We then confirm what was expected from the linear analysis, namely that this instability is relevant only for ultra-relativistic and even mildly relativistic shock precursors.

\subsubsection{Investigating the wave number dependence}
We have performed several simulations of upstream media of ultra-relativistic shocks having $\Gamma_{\rm sh}=100$  for various magneto-acoustic wavenumbers ranging from $\lambda=6.25\times 10^{-5}\ell_{\rm CR}$ to $\lambda=10^{-3}\ell_{\rm CR}$. The growth rates resulting from these simulations are displayed in Fig.(\ref{F6}). 
We once again see that the numerical results are consistent with the linear growth rate since the growth rate of the perturbation is increasing with the wave vector of the wave $k_x$. More precisely, we see that the $k_x^{1/2}$ dependence of the linear prediction is confirmed as the linear growth rates (displayed as dotted lines) fit the numerical growth rates well as long as the relative perturbation is small compared to unity. This numerical experiment confirms the theoretical prediction that these small scale waves are likely important contributors to magnetic turbulence in the shock precursor.\\
In order to evaluate the maximal growth rate expected from such instability, we have considered waves with the smallest wavelength possible, namely waves having $k_x\sim\ell_{\rm MHD}$ where $\ell_{\rm MHD}$ is the scale limit for the validity of MHD. The definition of $\ell_{\rm MHD}$ is $\beta_AR_{L,pl}$ where $\beta_a=V_a/c$ and $R_{L,pl}$ is the Larmor radius of the thermal protons. On the other hand the size of the precursor $\ell_{\rm CR}$ is of the order of the Larmor radius $R_{L*}$ of the most energetic CR generated by the shocks. The largest growth rate in an ultra-relativistic shock  will then be
\be
\gamma_x^{\rm max}\ell_{\rm CR} \simeq \xi_{\rm CR}\left(\frac{R_{L*}}{2R_{L,pl}}\right)^{1/2} \sim \xi_{\rm CR}\left(\frac{\Gamma_*}{2\Gamma_{\rm sh}\beta_a}\right)^{1/2}  
\ee
where $\Gamma_*$ is the Lorentz factor of the most energetic CRs generated by the shock. For standard values of density and magnetic field occurring in the ISM, one gets $\beta_a\sim 10^{-4}$. The previous growth rate is much larger than unity as long as 
\be
\xi_{\rm CR} \gg \left(\frac{10^{-4}\Gamma_{\rm sh}}{\Gamma_*}\right)^{1/2}
\ee
which for an ultra-relativistic shock $\Gamma_{\rm sh}=100$ and CRs with maximal energy $E_*=10^{15}EeV$ is equal to $\xi_{\rm CR}\gg10^{-4}$. It seems then most likely that efficient wave amplification may be achieved in the precursor of the shock. However it is too early to raise definite conclusions here, as only multi-dimensional calculations can catch the diversity of the source of free energy and provide a hint of the dominant MHD instability. We insist also on the issue that our work cannot provide any answer to the acceleration of high energy CRs in relativistic shocks yet, as again 3D calculations are mandatory for that purpose. The fact that the instability investigated here produces perturbations at scales $\lambda < r_L$ makes it necessary to also produce longer wavelength perturbations. Similar issues have been discussed in the context of supernova remnant shocks  \citep{Schure+12}.
\section{Summary and concluding remarks}
\label{Summary}
In this paper we have provided an analysis of the non-resonant streaming instability occurring in the precursor of astrophysical shock waves. Our investigation consists of a linear analysis complemented by numerical relativistic MHD simulations.  Two assumptions have been made to perform our work, namely that the MHD waves have a wavevector perpendicular to the shock front and a magnetic field to be parallel to the shock front. This configuration is likely to occur near ultra-relativistic shock fronts because of the properties of the Lorentz transformation. The non-resonant streaming instability is based on the presence of CRs in the precursor that lead to an electrically charged thermal plasma. The resulting electromotive force on the thermal plasma can destabilize MHD waves propagating through the precursor of the shock. \\ 

The outcome of our linear analysis is presented in Sect.(\ref{Linear}) and shows that under the aforementioned assumptions, magneto-acoustic waves will be destabilized by the electromotive force and can reach a saturation regime if its wave vector $k_x$ is large enough compared to the inverse of the size of the precursor $\ell_{\rm CR}$. The linear calculation predicts a $k_x^{1/2}$ dependence regarding the growth rate of the incoming wave. It also predicts that the shock velocity can influence the actual value of the growth rate as the growth rate is also proportional to $(\Gamma_{\rm sh}-1)/\Gamma_{\rm sh}\beta_{\rm sh}^{1/2}$. This factor has a maximal value of unity for ultra-relativistic shocks while it is decreasing with shock velocity $\beta_{\rm sh}$ as one enters the non-relativistic shock regime.  Ultra-relativistic shocks are then the best candidates for efficient non-resonant streaming instability. \\

We have performed numerical relativistic MHD simulations in order to follow the propagation of a magneto-acoustic wave through the precursor of a magnetized shock. Under the same framework than in the linear analysis, we were able to depict the whole propagation of the wave from its entrance in the precursor  to the non linear saturation regime. We have illustrated the basic stages of the instability and were able to compare the growth rate provided by numerical simulations with the computed linear growth rate. The numerical simulations match the growth rate predicted by linear theory and confirm that under our assumptions, non-resonant streaming instability is more efficient in the context of ultra-relativistic shocks than in the non-relativistic case. The growth rate of the instability is found to be proportional to $k_x^{1/2}$ ($k_x$ being the wavenumber of the perturbation).  Numerical simulations also highlight the non-linear growth of the instability in the final stages of the propagation of the wave. During that stage, the shape of the initial perturbation has been completely distorted while the amplitude of the magnetic perturbation has increased with $\delta B/B$ of the order of a few units. This stage corresponds to the saturation regime of the instability since the density perturbation is following the same behavior than the magnetic perturbation. Ultimately though, numerical errors occur when in the full saturation regime, we get true local vacua with $\delta\rho/\rho\rightarrow -1$. We found that the precursor of the shock can easily be populated by a train of shocks produced by the density and magnetic spikes. These shocks can actively participate to the local heating process of the background plasma as well as possibly to further particle acceleration and transport. This aspect will be investigated in a future work. We also confirmed that Alfv\'en disturbances are stable in this 1D flow configuration.\\

We restrict ourselves in this work to one-dimensional simulations and linear calculations of magneto-acoustic waves amplification. This choice was motivated by the natural configuration expected from an ultra-relativistic shock where the Lorentz transform leads to perpendicular magnetic fields and MHD waves having $k_x\gg k_y,k_z$. Compared to PLM09, we identified a new efficient instability regime 
that we believe is the most dominant mode among all instability regimes triggered by the presence of CR electrical charge density. Only considering multi-dimensional calculations will permit to isolate the dominant instability that can provide magnetic amplification. Also 3D simulations are necessary to account properly for the particle transport and derive the maximum CR energies in relativistic shocks. Hence we make it clear that the question of the dominant MHD instability and maximum CR energy in relativistic and mildly relativistic shocks can not be discussed within this work. A further step will be to deal with two-dimensional simulations of the system but this time taking into account the effect of the electrical current driven by the CR flow. This external current may be a promising element to the shock structure since it may have the ability to amplify Alfv\'en waves and thus provide higher  $\delta B/B$ ratios.

 \section*{Acknowledgements}
 F.C. and A.M. would like to thank G. Pelletier, M. Lemoine and I. Plotnikov for fruitful discussions. A.M. acknowledges the International Space Science Institute (ISSI) in Bern where preliminary stages of this work have been discussed during different sessions of a working group. Simulations presented in this paper were performed using HPC resources from GENCI-CINES (Grant 2012046325). R.K. acknowledges funding from FWO-Vlaanderen, project G.0238.12.

 \appendix
 
 \section{Isothermal SRMHD: switching from conservative to primitive variables}
 
 The calculations presented in this paper involves physical conditions where a large contrast between physical quantities occur. Indeed, in the context of ultra-relativistic MHD shocks, thermal pressure and magnetic pressure existing in the ISM are sensed as very small in the shock frame because of the Lorentz transformation. For instance, the simulation presented in Sect.(\ref{Sect:Figu}) exhibits thermal pressure $10^{12}$ times smaller than the kinetic energy density of matter. Dealing with such contrast is  tricky and requires a specific approach. In order to avoid having negative thermal pressure in our simulation we add an extra relation linking thermal pressure to density such that $P=\rho c_S^2$ where $c_S\geq 0$ is the constant sound speed in the shock frame.   
 Indeed, performing SRMHD simulations requires very frequent switches from conservative $(D,\bm{S},\tau,\bm{B})$ to primitive variables $(\rho,\bm{v},P,\bm{B})$. In order to achieve this transformation, it is convenient to use auxiliary variables such as the Lorentz factor $\Gamma$ or relativistic enthalpy $\xi=\Gamma^2(\rho c^2+\gamma P/(\gamma -1))$ (see \citet{vdHol08}). \\
 
 The major issue in switching variables comes from the fact that computing auxiliary variables leads to complex equations. In the case of  isothermal SRMHD, the two auxiliary variables read as
 \begin{eqnarray}
 \frac{1}{\Gamma^2} &=& 1 - \frac{(\bm{S}+\xi^{-1}(\bm{S}\cdot\bm{B})\bm{B})^2}{(\xi+\bm{B}^2)^2} \\
 \xi &=& \Gamma D\left(1 + \frac{\gamma}{\gamma -1}\frac{c_S^2}{c^2}\right)=\Gamma D\cal{F}
 \end{eqnarray}
 where  $\bm{B}^2$ denotes the magnetic pressure (magnetic permeability of vacuum set to unity) and $\cal{F}$ is a constant. Solving this systems leads to a fourth order polynomial in $\Gamma$
 \begin{eqnarray}
 &&\Gamma^4D^2{\cal F}^2 +2\Gamma^3D{\cal F}\bm{B}^2+\Gamma^2(\bm{B}^4-\bm{S}^2-D^2{\cal F}^2)- \\
 &&2\Gamma \left(\bm{B}^2D{\cal F} + 
 \frac{(\bm{S}\cdot\bm{B})^2}{D{\cal F}}\right) -\bm{B}^4- \frac{(\bm{S}\cdot\bm{B})^2\bm{B}^2}{D^2{\cal F}^2} = 0\nonumber
 \end{eqnarray}
 Hopefully, a single root of this polynomial corresponds to the actual value of the Lorentz factor. This is easy to show in the present framework as we can assume $\bm{B}^2\ll D$ and ${\cal F}\simeq 1$. The momentum then becomes $\bm{S}^2\simeq \Gamma_o^2D^2$ ($\Gamma_o$ being the root of the polynomial) and the polynomial can be written as 
 \begin{eqnarray}
 f(\Gamma) =\Gamma^4D^2 +2\Gamma^3D\bm{B}^2&-&\Gamma^2(\Gamma_o^2+1)D^2- \nonumber\\
 2\Gamma\left(\bm{B}^2D + \frac{(\bm{S}\cdot\bm{B})^2}{D}\right) &\simeq&\Gamma^4D^2-\Gamma^2(\Gamma_o^2+1)D^2
 \end{eqnarray}
 The relevant values of the Lorentz factor are such that $\Gamma\geq 1$ so we have to prove that the polynomial only has one root in the range $[1,+\infty[$. In that context, It is obvious that $f(\Gamma=1)\simeq -D^2<0$ while 
  \be
 \lim_{\Gamma\rightarrow\infty} f(\Gamma) = \Gamma^4D^2 > 0
 \ee
 The derivative of $f(\Gamma)$ is 
 \be
 \frac{df}{d\Gamma} \simeq 2\Gamma D^2\left(2\Gamma^2-\Gamma_o^2-1\right)
 \ee
 which is negative if $\Gamma\leq \Gamma_{\rm crit}=\sqrt{(\Gamma^2_o+1)/2}<\Gamma_o$ and positive otherwise. The function $f$ is then decreasing and remaining negative for $\Gamma \in \ [1,\Gamma_{\rm crit}] $ and strictly increasing for $\Gamma \in\ ]\Gamma_{\rm crit},+\infty[$. It is then obvious that only one root of the polynomial lies in the interval $[1,+\infty[$ as long as $\bm{B}^2,P\ll D$. Finding the root of function $f(\Gamma)$ is achieved using Newton-Raphson algorithm displaying a quartic efficiency (see \citet{vdHol08}).

\bibliographystyle{/sw/share/texmf-dist/tex/latex/aa/bibtex/aa}

\end{document}